\documentclass[journal]{IEEEtran}
\usepackage{amssymb}
\usepackage{amsmath,amsthm,amssymb}
\usepackage{graphicx}
\usepackage{epsf}
\usepackage{color}
\usepackage{algorithm}  
\usepackage{enumerate}
\usepackage[noend]{algorithmic}
\allowdisplaybreaks
\begin{document}
\raggedbottom
%\markboth{IEEE Transactions on Information Theory}{}
\sloppy
\newtheorem{claim}{Claim}
\newtheorem{corollary}{Corollary}
\newtheorem{definition}{Definition}
\newtheorem{example}{Example}
\newtheorem{exercise}{Exercise}
\newtheorem{fact}{Fact}
\newtheorem{lemma}{Lemma}
\newtheorem{note}{Note}
\newtheorem{obs}{Observation}
\newtheorem{problem}{Problem}
\newtheorem{property}{Property}
\newtheorem{proposition}{Proposition}
\newtheorem{question}{Question}
\newtheorem{ru}{Rule}
\newtheorem{solution}{Solution}
\newtheorem{remark}{Remark}
\newtheorem{theorem}{Theorem}

\newcommand{\A}{{\bf a}}
\newcommand{\CC}{{\cal C}}
\newcommand{\B}{{\bf b}}
\newcommand{\I}{{\bf i}}
\newcommand{\p}{{\bf p}}
\newcommand{\bh}{{\bf h}}
\newcommand{\q}{{\bf q}}
\newcommand{\x}{{\bf x}}
\newcommand{\y}{{\bf y}}
\newcommand{\z}{{\bf z}}
\newcommand{\M}{{\bf M}}
\def\Mt{\tilde{M}}
\newcommand{\N}{{\bf N}}
\newcommand{\X}{{\bf X}}
\def\r {{\bf r}}
\newcommand{\bP}{{\bf P}}
\newcommand{\Q}{{\bf Q}}
\newcommand{\U}{{\bf u}}
\def\d {{\tt d}}
\def\D {{\tt D}}
\def\V {{\tt V}}
\def\Av {{\tt Avg}}
\def\restr{\downharpoonright}
\newcommand{\comment}[1]{\ $[\![${\normalsize #1}$]\!]$ \ }

\newcommand{\restrict}{{\mathbin{\vert\mkern-0.5mu\grave{}}}}

\newcommand{\glb}{{\tt glb}}
\newcommand{\lub}{{\tt lub}}

\newcommand{\LL}{{\cal L}}
\newcommand{\PP}{{\cal P}}
\newcommand{\pv}{{{\bf p}=(p_1,\ldots , p_n)}}

\newcommand{\qv}{{{\bf q}=(q_1,\ldots , q_n)}}
\newcommand{\Inf}{{\bf t}}
\newcommand{\Sup}{{\bf s}}
\newcommand{\inff}{{\tt t}}
\newcommand{\supp}{{\tt s}}
\def\H {{\cal H}}
\def\l {{\bf l}}

\allowdisplaybreaks
\newcommand{\g}{{\bf g}}
\newcommand{\s}{{sup}}
\def\i {{inf}}
\def\r {{\bf r}}
\def\t {{\bf t}}
\def\m{{\bf m}}
\def\half{{\sf half}}
\newcommand{\pcoin}{{coin of bias $p$ }}
\newcommand{\ptree}{{tree of bias $p$ }}
\newcommand{\remove}[1]{}

\newcommand{\todo}[1]{{\tiny\color{red}$\tiny\circ$}%
{\marginpar{\flushleft\\scriptsizes\sf\color{red}$\bullet$#1}}}

\thispagestyle{plain}

\title{Minimum--Entropy Couplings and their Applications}
\author{Ferdinando Cicalese, Luisa Gargano, \IEEEmembership{Member, IEEE} and Ugo Vaccaro, \IEEEmembership{Senior Member, IEEE}%
\thanks{F. Cicalese is with the Dipartimento di Informatica,  Universit\`a di Verona, Verona, 
     Italy (email: ferdinando.cicalese@univr.it), L. Gargano is with the Dipartimento di Informatica,  Universit\`a di Salerno,
		Fisciano (SA), Italy (email: lgargano@unisa.it), 
 and U. Vaccaro is with the Dipartimento di Informatica,  Universit\`a di Salerno,
		Fisciano (SA), Italy (email: uvaccaro@unisa.it).	This paper was presented in part at the 2017 IEEE International Symposium on Information Theory.
			}}
\maketitle
\begin{abstract}
Given two discrete random variables $X$ and $Y,$ with probability distributions
$\p=(p_1, \ldots , p_n)$ and $\q=(q_1, \ldots , q_m)$, respectively,
denote by $\CC(\p, \q)$ 
the set of all 
couplings of $\p$ and $\q$, that is, the set of  all  bivariate probability distributions 
that have $\p$ and $\q$ as marginals.
In this paper, we study  the problem of
finding a joint probability distribution in $\CC(\p, \q)$ of
\emph{minimum entropy} (equivalently, a coupling
that \emph{maximizes} the mutual information between $X$ and $Y$),
and we discuss several situations where the need for
this kind of optimization  naturally arises. 
Since the optimization problem is known to be NP-hard,
we give an efficient algorithm to find a joint probability distribution in $\CC(\p, \q)$
with   entropy  exceeding the minimum possible at most by {1 bit}, 
thus providing an approximation algorithm with an additive gap of at most 1 bit. 
Leveraging on this algorithm, we extend our result to
the problem of finding a minimum--entropy joint distribution
of arbitrary $k\geq 2$ discrete random variables $X_1, \ldots , X_k$, consistent
with the known $k$ marginal distributions of the individual random variables $X_1, \ldots , X_k$.
In this case, our algorithm has an { additive gap of at most $\log k$ from optimum.} 
 We also discuss several  related  applications  of our findings and  {extensions of
our results to entropies different from the Shannon entropy.}
\end{abstract}

\section{Introduction and Motivations}
Inferring  an unknown joint distribution of two random variables (r.v.),
 when only their  marginals are given, is an old problem 
in the area of probabilistic inference. The problem goes back at least to Hoeffding \cite{Hoeff}
and Frechet \cite{frechet},
who studied the question of identifying the extremal joint distribution of
r.v. $X$ and $Y$ that maximizes (resp., minimizes) their correlation, 
given the    marginal distributions of $X$ and $Y$.
We refer the reader to \cite{benes,cuadras,Aglio,Lin} for a (partial) account of the
 vast literature in the area and the many  applications in the pure and applied sciences.

In this paper, we consider the following  case  of the   problem  described above. Let $X$ and $Y$ be 
two discrete r.v., distributed according to $\p=(p_1, \ldots , p_n)$ and $\q=(q_1, \ldots , q_m)$, respectively. 
We seek a \emph{minimum-entropy} joint probability distribution of $X$ and $Y$, whose   marginals are equal  to 
$\p$ and $\q$. 

We discuss below a few scenarios where this problem naturally arises.

\subsection{Entropic Causal Inference}
In papers \cite{ko1} and \cite{ko2},  the authors 
consider the  important question  of identifying the correct causal direction between two arbitrary
r.v.'s $X$ and $Y$, that is,  they want to discover whether it is  the case that $X$ 
causes $Y$ or  it is  $Y$  that causes $X$.  In general,
$X$  causes $Y$ if there exists an exogenous  r.v. $E$ (independent { of} $X$) and a deterministic function $f$ such that
$Y=f(X{,} E)$. In order to identify the correct causal direction (i.e., either from $X$ to $Y$ or 
from $Y$ to $X$), the authors of \cite{ko1} { and} \cite{ko2}  make the reasonable postulate that the entropy of the  exogenous  r.v. $E$ is 
{small} in the 
\emph{true} causal direction, and empirically validate this assumption.
Additionally, they prove the important  fact that 
the problem of finding the exogenous variable $E$  with minimum entropy
 is equivalent to the problem of finding the minimum-entropy joint
distribution of properly defined random variables, \emph{given} (i.e., { by fixing}) their marginal distributions
({see} Theorem 3 of \cite{ko2}).
This is exactly    the problem we consider in this paper.
The authors of \cite{ko2}  observe that the latter optimization 
problem is NP-hard (due to results of \cite{KSS15} and \cite{V12}) and 
 propose  
a greedy  \emph{approximation} algorithm  to find the minimum-entropy joint
distribution, given the marginals.  For this greedy  algorithm, 
the authors prove  that
it  always finds a local minimum and that the local minimum is 
within an {  additive guaranteed gap} 
from the unknown global optimum. The authors of \cite{ko2} observe that this 
{ additive guaranteed gap} 
can be as large as $\log n$ (here $n$ is the cardinality of the support of each  involved random variable).
Similar  results are  contained in \cite{Pai},  and references therein.

In this paper, we design  a different 
 greedy algorithm, 
and   we  prove that it
returns a correct  joint probability distribution (i.e., with 
the prescribed marginals)
with   entropy  exceeding the \emph{minimum possible by at most { 1 bit}}. 
Subsequently, in Section \ref{sec:more}, we extend our algorithm to the case of more than two
random variables.  More precisely, 
we consider 
the problem of finding a minimum--entropy joint distribution
of arbitrary $k\geq 2$ discrete random variables $X_1, \ldots , X_k$, consistent
with the known $k$ marginal distributions of $X_1, \ldots , X_k$.
In this case, our algorithm has an additive {guaranteed gap of at most}
$\log k$. 

{\subsection{On the functional representation of correlated random variables}\label{sub:fl}
Let $X$ and $Y$ be two arbitrary random variables with joint distribution $p(x,y)$.
The functional representation lemma \cite[p. 626]{Gamalbook} states that 
 there exists a random variable $Z$ independent
{of}  $X$, and a {deterministic} function $f$,  such that the pair of 
r.v.{'s} $(X,f(X,Z))$ is distributed like  $(X,Y)$, 
that is, they have the same  joint distribution $p(x,y)$.
This lemma has been applied  to establish several results in network information
theory (see \cite{Gamalbook} and references therein).
In several applications, it is important to find { a} r.v. $Z$ such that 
the { conditional entropy} $H(Y|Z)$ is close to its natural lower bound, that is, it is close to
$I(X;Y)$. Recently, a very strong result to that respect was proved in 
\cite{LiGamal}, showing that one can indeed find a r.v.\ $Z$ such that
$H(Y|Z)\leq I(X;Y)+\log (I(X;Y)+1)+4$ { bits}. Among the papers 
that have used (versions of) the functional representation lemma,
{  papers \cite{Bra} and \cite{Har}} have considered the problem
of one-shot channel simulation with unlimited common randomness.\footnote{The situation is also 
somewhat reminiscent of the important ``reverse Shannon Theorem'' 
 of \cite{Benn}, where one wants to simulate an arbitrary noisy channel 
with a noiseless one, plus some   additional source of randomness (see \cite{Benn} for formal definitions).} 
The bounds on $H(Y|Z)$ are essentially used to set a limit on the amount
of bits exchanged among parties involved in the simulation. 
However, there is here another resource { which is} reasonable
to bound: the amount of randomness used in the protocol, { i.e.,},
the amount of randomness needed to generate the auxiliary r.v. $Z$. 
Indeed,  randomness \emph{is not} free, and several clever (but expensive) methods 
have been devised to produce it, based on physical 
systems like Geiger-Muller tubes,  chaotic laser, etc.;
therefore it seems reasonable to require that the entropy of auxiliary   r.v. $Z$
be \emph{minimum}\footnote{This requirement can be made more formal
by invoking the fundamental result of Knuth and Yao \cite{KY}, stating  that the
minimum average number of unbiased random bits necessary to generate an arbitrary discrete  r.v. $Z$ is 
sandwiched between $H(Z)$ and $H(Z)+2$.}.  
On the basis of the (already mentioned)  important  result proved in \cite{ko1}, showing the equivalence between   
the problem of finding the minimum entropy auxiliary  variable $Z$  such that $(X,f(X,Z))=(X,Y)$,
and  the problem of finding the minimum-entropy joint
distribution of properly defined random variables (given their marginal distributions),
it follows that the results of our  paper offer a solution to the question
of seeking a minimum entropy r.v. $Z$ such that the pair of r.v. $(X;Y)$ can be 
simulated as $(X,f(X,Z))$. The exact statement of our result { is} given in Corollary  \ref{cor:frl}
of  Section \ref{sec:more}.
% when all the necessary notations have been explained.}

\subsection{Metric for dimension reduction}
Another work  that considers the problem of finding the 
minimum-entropy joint  distribution 
of two r.v. $X$ and $Y$, given the marginals of $X$ and $Y$, is the paper 
\cite{V12}. There, the author introduces   a pseudo-metric $\D(\cdot, \cdot)$ among discrete 
probability distributions in the following way:
given arbitrary $\p=(p_1, \ldots, p_n)$ 
and $\q=(q_1, \ldots , q_m)$,  the distance $\D(\p,\q) $
among $\p$ and $\q$  is defined as the quantity
 $$\D(\p,\q)=2W(\p,\q)-H(\p)-H(\q),$$
where $W(\p,\q)$ is the \emph{minimum} entropy of a bivariate
probability distribution that has $\p$ and $\q$ as marginals,
and $H$ denotes the Shannon entropy. This metric is applied  in \cite{V12}  to the problem of dimension-reduction
of stochastic processes.
The author of \cite{V12} observes that the problem of computing
$W(\p,\q)$ is NP-hard (see also \cite{KSS15}) and proposes another different  greedy algorithm for its
computation, based on some analogy with the problem of Bin Packing
with overstuffing. No performance guarantee is given in
\cite{V12} for the proposed algorithm. Our result directly implies
that we can compute the value of the pseudometric $\D(\p,\q)$, for \emph{arbitrary} $\p$ and $\q$,
with an additive { gap} of at most $1$ { bit}.\footnote{We remark that in \cite{CGVold,CGVnew}
we considered   the different problem of computing  the
probability distributions $\q^*$  that \emph{minimizes} $\D(\p,\q)$,
given $\p$.}

\subsection{Contingency tables and transportation polytopes}
In the field of Combinatorial Optimization, the set $\CC(\p,\q)$ of all 
couplings of given $\p=(p_1, \ldots, p_n)$ and $\q=(q_1, \ldots , q_m)$  is known as the transportation 
polytope $P$ defined by $\p$ and $\q$. The fact that in our case
$\p$ and $\q$ are probability distributions (i.e., their components { are non-negative and} sum up
to 1) is immaterial, since one can always normalize. 
A similar concepts is known in Statistics under the name of  contingency tables \cite{Do}.
Polytopes $\CC(\p,\q)$
are  called  transportation polytopes 
 because they model the transportation of goods
from $n$  supply locations (with the $i$-th location supplying a quantity of $p_i$) to $m$
demand locations (with the $j$-th location demanding a quantity of $q_j$). The feasible
points $M_{i,j}$ of an element $\M=[M_{i,j}]\in \CC(\p,\q)$ model the scenario
where a quantity of $M_{i,j}$ of goods is transported from the $i$-th supply location to the
$j$-th demand location. Many hard combinatorial optimization problems
become solvable on the transportation polytope because of its rich and well studied 
combinatorial structure. We refer to the survey paper \cite{De}
for an account of the vast  literature on the topic. 
The problem we consider in this paper can be equivalently stated as the one of  finding a minimum-entropy 
element of 
 the polytope $\CC(\p,\q)$. To see that this is not  a simple translation of 
a problem from one language into another, we  point out
a recent important  trend in the area of combinatorial optimization, that is, the one
that seeks sparse solutions to optimization problems.
More precisely,  researchers aim at  finding algorithms  that trade
the optimality of a solution to a given problem with the sparseness 
of the solution (e.g., the number of variables with non-zero values
in the solution, but other measure of sparseness can be employed).
 We address the reader to  \cite{Ba1,shai}, and references therein, for motivations  and 
a review
 of this line of research. 
Our problem of finding a minimum entropy element in the transportation polytope  $\CC(\p,\q)$
fits in the above perspective.  This interpretation is possible not  only because entropy can be often 
interpreted as a reasonable  measure of sparseness (see \cite{hu})
but also because our algorithm   produces  an elements 
of $\CC(\p,\q)$ whose number of non zero elements is, in the worst case,  at most 
twice the minimum possible. We remark that finding a matrix  $\M\in \CC(\p,\q)$  with the 
minimum 
number of non-zero entries  is  NP-hard  in general \cite{ko1}.

\subsection{Additional relations and applications}
We believe  that the problem of finding a minimum entropy joint
distribution, with fixed marginal, is indeed a basic  one. In this section we
will briefly illustrate a few other scenarios where the problem matters.

\medskip

Let us write the joint entropy of two r.v. $X$ and $Y$,
distributed according to $\p$ and $\q$, respectively,  as
$H(X,Y)=H(X)+H(Y)-I(X;Y)$, where $I(X;Y)$ is the mutual information
between $X$ and $Y$. Then,  one sees that our original problem 
can  be equivalently stated as the determination  of  a joint probability distribution of
$X$ and $Y$ (having   given marginals $\p$ and $\q$) that \emph{maximizes} the mutual information
$I(X;Y)$. In  \cite{KSS15}, this maximal mutual information is interpreted,
in agreement with Renyi's axioms for a \emph{bona fide} dependence measure \cite{renyi},
as a measure of the \emph{largest possible} correlations between  two r.v. $X$ and $Y$.
One can see the soundness of this   interpretation also in the following way. 
Let $q(x,y)$ be an arbitrary joint distribution of r.v. $X$ and $Y$, with 
marginals equal to $\p=\{p(x)\}$ and $\q=\{q(y)\}$, respectively. Then,
it is well known that the mutual information $I(X;Y)$ can be written
as the relative entropy (divergence) between the joint distribution $q(x,y)$ and
the joint distribution $r(x,y)=p(x)q(y)$ (i.e., a joint distribution that would 
make $X$ and $Y$ \emph{independent}). Therefore, our problem of maximizing $I(X;Y)$  is equivalent 
to the one of finding a joint distribution 
of $X$ and $Y$ that is the \emph{farthest}
(in the sense of relative entropy) from $r(x,y)=p(x)q(y)$, that is, finding  
the joint distribution of $X$ and $Y$ that makes them ``most dependent'' (or correlated) 
as possible. 
Another way to see the question is to realize that we are seeking a
joint distribution that minimizes the conditional entropies $H(X|Y)$ and $H(Y|X)$, that 
represent sensible measures of the strength of the dependence between $X$ and $Y$.
  Since the problem of its exact computation is  NP-hard, our result
implies  an approximation algorithm for it. We would like to remark that there
are indeed situations in which measuring the ``potential'' correlation between 
variables (as opposed to their actual correlation) can be  useful. For instance, the 
authors of \cite{Kri} introduces a measure 
that, in their words,  ``\emph{provides a score for the strength of the influence protein X has on protein Y. 
In many
physiological conditions, only a small fraction of the cells have activated protein X in response
to stimuli, and these active populations have little influence on the mutual information metric}''.
Since other standard measures of correlation would also fail,  using a measure of \emph{potential}
correlation in that context 
could be useful.

Another situation  where  the need to maximize the mutual information 
between two r.v. (with fixed probability distributions) naturally   arises, is in the area of medical imaging
 \cite{pluim, wells}.

In   \cite{Rou}, the authors asks several algorithmic problems of this vein: given some 
fixed marginal distributions, find a  joint probability distribution,  
  with  marginals \emph{close} to the given ones, and 
 { satisfying} some additional properties  dictated by certain applications scenarios.
Among the several problems considered in \cite{Rou}, the authors  mention the problem 
of finding a minimum entropy joint distribution, with  the given marginals, as an interesting
open problem.

In the recent paper \cite{Yu}, the authors study the problem of finding a  joint probability distribution of two 
random  variables with fixed marginals,  that minimizes a given function $f$ of the joint distribution. 
The authors study this problem in the \emph{asymptotic setting} (i.e., for product marginal distributions).
A strictly related problem was also studied in \cite{verdu}.

In \cite{Han}, the authors study the problem of finding good upper and lower
bounds on the mutual information $I(X;Y)$ of two r.v.{'s} $X$ and $Y$ when the only 
available knowledge consists of   the marginals of $X$ and $Y$, and the pair of values
$(x,y)$ for which the unknown joint distribution of $X$ and $Y$ (consistent with the given 
marginals) assign a non-zero probability. It is clear that our maximization problem
 gives an upper bound on $I(X;Y)$ when the 
available knowledge consists of   the marginals of $X$ { and} $Y$, and \emph{nothing else}.

Other papers  considering problems somewhat related to ours are \cite{fritz,Kap, Maes, Miller,Sa, Whitt}, and \cite{Yuan}.

\subsection{Structure of the paper}

The rest of the paper is organized as follows. 
In Section \ref{sec:pre} we present  the mathematical
tools and the auxiliary results that are needed to prove our results.
In Section \ref{sec:approx} we present our algorithm to find a joint 
probability distribution of two input r.v.{'s} $X$ and $Y$, with given 
marginal distributions, whose entropy is at most one bit away from the 
joint distribution of minimal entropy. We also present a worked out example 
to illustrate the behaviour of the algorithm in an intuitive way. 
The formal proofs of the correctness of the algorithm are 
spelled out in the successive Section \ref{sec:proof}.
In Section \ref{sec:more} we extend the algorithm presented in Section \ref{sec:approx}
to an arbitrary number of $k\geq 2$ input random variables.
The entropy of the joint distribution produced by our algorithm is at most $\log k$  bits away from the 
minimum-entropy joint distribution of the $k$ r.v.'s. 

Throughout this paper, the logarithms are on base 2 unless specified otherwise.
\section{Preliminary Results}\label{sec:pre}

To prove our results, we  use ideas  and techniques  from majorization theory \cite{MO}, 
a mathematical framework that has been  proved to be very  much useful  in 
information theory (e.g., see \cite{CV,CV1,CGV,CGVold,Erven,HY,HV,Sa2} and references therein).
In this section we recall the notions and results 
that  are relevant to our context.

\begin{definition}\label{defmaj} 
Given two probability distributions
$\A=(a_1, \ldots ,a_n)$ and $\B=(b_1, \ldots , b_n)$ with $a_1\geq \ldots \geq a_n\geq 0$ and 
$b_1\geq \ldots \geq b_n\geq 0$, $\sum_{i=1}^na_i=\sum_{i=1}^nb_i=1$, we say that $\A$ is 
{\em majorized} by $\B$, and write  $\A \preceq \B$,
if and only if
$\sum_{k=1}^i a_k\leq \sum_{k=1}^i b_k, \quad\mbox{\rm for all }\  i=1,\ldots , n.$
\end{definition}
We assume that {all the probability distributions 
we deal with have been
ordered in non-increasing order}. This assumption does not affect our results, since the quantities we compute
(i.e., entropies) are invariant with respect to permutations of the components of
the involved probability distributions. We also  use the majorization 
relationship between vectors of unequal lengths, by properly padding the shorter
one with the appropriate number of $0$'s at the end.
The majorization relation $\preceq$ is a partial ordering on the $(n-1)$-dimensional simplex
$$\PP_n=\{(p_1,\ldots, p_n)\ :  \sum_{i=1}^n p_i=1, \ p_1\geq \ldots \geq p_n\geq 0\}$$  
of all ordered probability vectors of $n$ elements, that is, for each $\x,\y,\z\in\PP_n$
it holds that 
\begin{itemize}
\item[1)] $\x\preceq \x$;
\item[2)] $\x\preceq \y$ and $\y\preceq \z$  implies $\x\preceq \z$;
\item[3)]  $\x\preceq \y$ and $\y\preceq \x$  implies $\x =\y$.
\end{itemize}

It turns out that the partially ordered set $(\PP_n,\preceq)$ is indeed a \emph{lattice} \cite{CV},\footnote{The same result was independently 
rediscovered in \cite{cuff}, see also \cite{harr} for a different   proof.} 
i.e., for all $\x,\y  \in \PP_n$
there exists
a unique
{\em least upper bound}  $\x\lor \y$ and 
 a unique {\em greatest lower bound}  $\x \land \y$.
We recall that  the least upper bound
 $\x \lor \y$ is the vector in $\PP_n$ such that  
$$\x\preceq \x \lor \y, \  \y\preceq \x \lor \y,$$
   and 
for all  $\z\in \PP_n$ for which
$\x \preceq \z,\ \y \preceq \z$  it holds that  
$$\x\lor \y\preceq \z.$$
Analogously, the greatest lower bound $\x \land \y$ is the vector in $\PP_n$ such that
$$ \x \land \y\preceq \x,  \ \x \land \y\preceq \y,$$  
and 
for all $\z\in \PP_n$ for which 
$\z \preceq \x,\ \z \preceq \y$ it holds that 
$$\z\preceq \x\land \y.$$
In the  paper \cite{CV} the authors also gave a simple and efficient algorithm to explicitly 
compute $\x \lor \y$ and $\x\land \y$, given arbitrary vectors $\x,\y\in \PP_n$. 
Due to the important  role it will play in our main result, 
we recall how to  compute  the greatest lower bound. 
\begin{fact} \label{glb} {\rm\cite{CV}} 
Let $\x = (x_1, \dots, x_n), \y = (y_1, \dots, y_n) \in \PP_n$ and let $\z = (z_1, \dots, z_n) = \x \land \y$. 
Then, {$z_1=\min\{p_1, q_1\}$ and 
for each $i = 2, \dots, n,$ it holds that 
$$z_i = \min \Bigl \{\sum_{j=1}^i p_j, \sum_{j=1}^i q_j\Bigr\} 
- \sum_{j=1}^{i-1} z_{j}.$$  }

Equivalently, we have 
$$\sum_{k=1}^{i} z_k = \min \Bigl\{\sum_{k=1}^{i} p_k , \sum_{k=1}^{i} q_k  \Bigr\}.$$
Moreover, using 
$\sum_k z_k = \sum_k p_k = \sum_k q_k = 1,$ we also have 
that for each $i = 1, \dots, n,$ it holds that 
\begin{equation} \label{z-suffix-sum}
\sum_{k=i}^{n} z_k = \max \Bigl\{\sum_{k=i}^{n} p_k , \sum_{k=i}^{n} q_k  \Bigr\}.
\end{equation}
\end{fact}

\medskip
\noindent
We also remind   the important  Schur-concavity property  of the entropy function \cite{MO}:

\medskip
\emph{For any  $\x,\y\in\PP_n$, 
 $\x\preceq \y$ implies that    $H(\x)\geq H(\y)$, with equality 
    if and only if   $ \x=\y$. }
		
		\medskip
A notable strengthening  of above fact has been proved  in \cite{HV}. There, the authors prove that 
$\x\preceq \y$ implies 
\begin{equation}\label{eq:HV}
H(\x)\geq H(\y)+D(\y||\x),
\end{equation}
 where $D(\y||\x)$ is the relative entropy
between $\x$ and $\y$.

We also need the  concept of \emph{aggregation} { (see \cite{V12} and \cite{CGVold})}, and a result from \cite{CGVold},
whose proof is repeated here to make the paper self-contained.
Given $\p=(p_1, \ldots , p_n)\in \PP_n $ and an integer $2\leq m<n$,
we say that $\q=(q_1, \ldots , q_m)\in \PP_m$ is an \emph{aggregation} 
of $\p$ if there is a partition of $\{1, \ldots , n\}$ into disjoint sets $I_1, \ldots , I_m$
such that $q_j=\sum_{i\in I_j}p_i$, for $j=1, \ldots m$.

\begin{lemma}{\rm\cite{CGVold}}\label{pprecq}
Let $\q\in \PP_m $ be \emph{any} aggregation of $\p\in \PP_n$.  Then it holds that
$\p\preceq \q$.
\end{lemma}
\begin{IEEEproof} 
We shall prove  by induction on $i$ that $\sum_{k=1}^{i}q_k\geq \sum_{k=1}^{i}p_k$.
Because  $\q$ is an aggregation of $\p$, we know that there exists    $I_j\subseteq \{1, \ldots , n\}$
such that $1\in I_j$. This implies that  $q_1\geq q_j\geq p_1$. Let us suppose that 
$\sum_{k=1}^{i-1}q_k\geq \sum_{k=1}^{i-1}p_k$. 
If there exist  indices $j\geq i$ and $\ell\leq i$ such that $\ell\in I_j$,
then $q_i\geq q_j\geq p_\ell\geq p_i$,  implying 
$\sum_{k=1}^{i}q_k\geq \sum_{k=1}^{i}p_k$.  
Should it be otherwise,  for each $j\geq i$ and $\ell\leq i$ it holds that $\ell\not \in I_j$. Therefore,
$\{1, \ldots ,i\}\subseteq I_1\cup \ldots \cup I_{i-1}$. This immediately gives
$\sum_{k=1}^{i-1}q_k\geq \sum_{k=1}^{i}p_k$, from which we 
get $\sum_{k=1}^{i}q_k\geq \sum_{k=1}^{i}p_k$.
\end{IEEEproof}

\bigskip

Let us now discuss some  first consequences of { the} above framework.
Given two discrete random variables $X$ and $Y$, with probability distributions
$\p=(p_1, \ldots , p_n)$ and $\q=(q_1, \ldots , q_m)$, respectively,
denote by $\CC(\p, \q)$ 
the set of all joint distributions of $X$ and $Y$ 
that have $\p$ and $\q$ as marginals. In the literature, elements of $\CC(\p, \q)$ are often called \emph{couplings} of 
$\p$ and $\q$, and play an important role in many information theoretic problems, e.g, see \cite{Sa}. 
For our purposes,
each element in $\CC(\p, \q)$ can be seen as { an} $n\times m$ matrix $M=[m_{ij}]\in \mathbb{R}^{n\times m}$
such that its row-sums give the elements of $\p=(p_1, \ldots , p_n)$ and its column-sums give the elements of $\q=(q_1, \ldots , q_m)$, that is, 
\begin{equation}\label{CC}
\CC(\p, \q)=\Bigl\{\mathbf{M}=[m_{ij}]: \sum_{j}m_{ij}=p_i, \sum_{i}m_{ij}=q_j\Bigr\}.
\end{equation}
Now, for any  $\mathbf{M}\in \CC(\p, \q)$, let us write its elements in a $1\times mn$ vector
$\mathbf{m}\in \PP_{mn}$, with its components ordered in non-increasing fashion.
From (\ref{CC}) we obtain  that both  $\p$ and $\q$ are  {aggregations}
of \emph{each } $\mathbf{m}\in \PP_{mn}$  obtained from some  $\mathbf{M}\in \CC(\p, \q)$. 
By Lemma \ref{pprecq}, we get that\footnote{Recall that we 
use the majorization 
relationship between vectors of unequal lengths, by properly padding the shorter
one with the appropriate number of $0$'s at the end. This trick does not affect our 
subsequent results, since 
we use the customary assumption that $0\log 0=0$.}
\begin{equation}\label{m<peq}
\mathbf{m}\preceq \p \quad \mbox{and} \quad \mathbf{m}\preceq \q.
\end{equation}
Recalling the definition and properties of the greatest lower bound
of two vectors in $\PP_{mn}$, we also obtain
\begin{equation}\label{m<pandq}
\mathbf{m}\preceq\p\land \q.
\end{equation}
From (\ref{m<pandq}), and the Schur-concavity of the Shannon entropy, we also obtain that
$$H(\mathbf{m})\geq H(\p\land \q).$$
Since, obviously, the entropy of $H(\mathbf{m})$ is equal to the entropy $H(\mathbf{M})$,
where $\mathbf{M}$ is the matrix in $\CC(\p, \q)$ from which the vector $\mathbf{m}$
was obtained, we get the following   result ({ a key one} for our purposes).
\begin{lemma}\label{lemma:HM>Hpandq}
\emph{For any} $\p$ and $\q$, and  $\mathbf{M}\in \CC(\p, \q)$, it holds that 
\begin{equation}\label{eq:HM>Hpandq}
H(\mathbf{M})\geq H(\p\land \q).
\end{equation}
\end{lemma}
Lemma \ref{lemma:HM>Hpandq} obviously implies that the minimum-entropy 
coupling of $\p$ and $\q$ that we are seeking satisfies the inequality
$$\min_{\mathbf{N}\in \CC(\p, \q)} H(\mathbf{N}) \geq H(\p\land \q),$$
and it is one of the  key results towards our algorithm
to find an  element $\mathbf{M}\in \CC(\p, \q)$ with entropy
{ at most $1$ bit larger than the entropy of the minimum entropy coupling, i.e.,}
$H(\mathbf{M})\leq OPT+1$, where 
$OPT=\min_{\mathbf{N}\in \CC(\p, \q)} H(\mathbf{N}).$

\subsection{An interlude}
Before describing our algorithm and its analysis, let us illustrate some  
consequences of Lemma \ref{lemma:HM>Hpandq} not  directly  aimed 
towards proving our main results, but nevertheless of some interest.

It is well known that for any joint distribution of the two r.v. $X$ and $Y$ it
holds that 
\begin{equation}\label{trivial}
H(X{,} Y)\geq \max\{H(X), H(Y)\}.
\end{equation}
Since $H(XY)=H(\mathbf{M})$, for some $\mathbf{M}\in \CC(\p, \q)$ (where $\p$ and $\q$ are
the marginal distributions of $X$ and $Y$, respectively), we   can formulate the bound (\ref{trivial})
in the following 
\emph{equivalent} way:

for any $\mathbf{M}\in \CC(\p, \q)$ it
holds that 
$$H(\mathbf{M})\geq \max\{H(\p), H(\q)\}.$$
Lemma \ref{lemma:HM>Hpandq} allows us to 
strengthen   the lower bound 
(\ref{trivial}). Indeed,  by the definition of the greatest lower bound $\p\land \q$ of 
probability distributions $\p$ and $\q$,
it holds that
$\p\land \q\preceq \p$ and $\p\land \q\preceq \q$, and therefore,  by the Schur-concavity of
the entropy function and Lemma \ref{lemma:HM>Hpandq} we get  the improved lower bound
\begin{equation}\label{improved}
H(XY)=H(\mathbf{M})\geq H(\p\land \q)\geq \max\{H(\p), H(\q)\}.
\end{equation}

Inequality (\ref{improved}) also allows us to improve on  the 
classical  upper bound on the mutual information given by $I(X;Y)\leq \min\{H(X), H(Y)\},$
since (\ref{improved}) implies
\begin{equation}\label{improved2}
I(X;Y)\leq H(\p)+H(\q)-H(\p\land \q)\leq \min\{H(X), H(Y)\}.
\end{equation}
The new bounds (\ref{improved}) and (\ref{improved2}) are \emph{strictly} better than the usual ones, whenever
$\p\not\preceq\q$ and $\q\not\preceq\p$.
Technically, one could 
improve them even more, by using the 
inequality $H(\x)\geq H(\y)+D(\y||\x)$, whenever $\x\preceq \y$
 \cite{HV}. However,   in this paper we just need what we can get from 
 the inequality  $H(\x)\geq H(\y)$, if   $\x\preceq \y$ holds.
 
Inequalities (\ref{improved}) and (\ref{improved2}) could be useful also in other contexts,  when
one needs to bound the joint entropy (or the mutual information) of two
r.v.'s $X$ and $Y$,
and the only available knowledge is given by  the marginal distributions of $X$ and $Y$
(and not their  joint distribution).
Let $X$ and $Y$ be two r.v.{'s}, where $X$ is distributed according
to $\p$ and $Y$ according to $\q$, and let $H(X|Y)$ be the conditional
entropy of $X$ given $Y$. From (\ref{improved2}) we get
\begin{align*}
H(X|Y)&=H(X)-I(X;Y)\\
      &\geq H(X)-H(\p)-H(\q)+H(\p\land \q)\\
			&=H(\p\land \q)-H(\q).
\end{align*}
The  inequality $H(X|Y)\geq H(\p\land \q)-H(\q)$ gives a lower bound on $H(X|Y)$ that does not
depend on the joint distribution of $X$ and $Y$. In particular, it also 
 implies that if the probability distributions $\p$ and $\q$ of $X$ and $Y$
are such that $\q \not\preceq\p$, then the   conditional entropy $H(X|Y)$
cannot be zero, \emph{no matter} what the joint distribution of $X$ and $Y$ is.
By the Fano inequality, one gets a lower bound of the error  probability $\Pr\{X\neq Y\}$
 that depends only on the ``structure'' 
of the probability distributions $\p$ and $\q$ of $X$ and $Y$ 
and not on the joint distribution of $X$ and $Y$.
Admittedly, this lower bound is weak, but the only fact that one could
derive one  that is independent from the joint distribution of $X$ and $Y$ seems novel and interesting to us.

\medskip
Another possible application of the framework of Section \ref{sec:pre}  concerns  the problem  of sumset estimates for  Shannon entropy
\cite{ko,Tao}. There, one wants to find upper and lower bounds on the entropy of $H(X+Y), H(X-Y)$ (and similar 
expressions), in terms of the individual entropies $H(X), H(Y)$. As an example, one could somewhat improve the trivial 
estimate $H(X)+H(Y)\geq H(X+Y)$ in the following way. 
Let us consider  $X+Y$ and observe that the probability mass function of $X+Y$ is an aggregation of the 
pmf of the joint random variable $(X,Y)$. Then, by  Lemma \ref{pprecq} and formula (\ref{eq:HV}), one immediately gets
the  inequality 
\begin{align}
H(X)+H(Y)&\geq H(X,Y)\geq H(X+Y)\nonumber\\
         &+D(X+Y||(X,Y)) \geq H(X+Y),\label{eq:ko1}
\end{align}
where the last inequality is strict unless the pmf of $X+Y$ is equal to that of $(X,Y)$.
Similar  improvements can be obtained for other expressions like $X-Y$. 
More in general, one has  the following inequality that holds for any determinist function $f$ and discrete
r.v. $Z$:
\begin{equation}\label{eq:ko2}
H(Z)\geq H(f(Z))+D(f(Z)||Z),
\end{equation}
where one recovers (\ref{eq:ko1}) when $Z=(X,Y)$ and $f(X,Y)=X+Y$.

\section{An Algorithm to Approximate   \ $OPT=\min_{\mathbf{N}\in \CC(\p, \q)} H(\mathbf{N}).$}\label{sec:approx}
In this section we present  our main result, that is, an algorithm that from the  input distributions
$\p$ and $\q$, constructs a coupling  $\mathbf{M}\in \CC(\p, \q)$ such that 
\begin{equation}\label{eq:HM<Hpandq+1}
H(\mathbf{M})\leq H(\p\land \q)+1.
\end{equation}
 Lemma \ref{lemma:HM>Hpandq} will imply  our desired result, that is
$$ H(\mathbf{M})\leq \min_{\mathbf{N}\in \CC(\p, \q)} H(\mathbf{N})+1. $$

The following lemma is technical in nature, but it turns out to be a very useful tool of our main algorithm.
\begin{lemma} \label{lemma:pezzettini}
Let $A[1 \dots k]$ be an array of $k$ 
non-negative real numbers and $z$ a positive real number such that $z \geq A[i]$ for each $i =1,\dots,k.$ 
For any $x \geq 0$ such that $x \leq z + \sum_{i=1}^k A[i]$  there exists a subset $I \subseteq \{1,\dots k\}$ 
and $0 \leq z^{(d)} \leq z$ such that  $$z^{(d)} + \sum_{i  \in I} A[i]  = x.$$ Moreover, $I$ and $z^{(d)}$ can 
be computed in linear time. 
\end{lemma}
\begin{IEEEproof}

If $\sum_{i =1}^k A[i] < x,$  the desired result is given by setting 
$I = \{1,\dots, k\}$ and $z^{(d)} = x - \sum_{i=1}^k A[i]$ which is a positive number 
not larger than $z$, from the assumption 
that $z + \sum_{i=1}^k A[i] \geq x.$ Note that 
the condition can be checked in linear time.

Let us now assume that $\sum_{i=1}^k A[i] \geq x.$
Let $j$ be the minimum index such that
$\sum_{i=1}^j A[j] \geq x.$ Then setting $I = \{1, \dots, j-1\}$  (if $j = 1,$ we set $I = \emptyset$) 
and---using the assumption that $z \geq A[j]$---$z^{(d)} = x - \sum_{i=1}^j A[j]$
we have the desired result. Note that also in this case the index $j$ can be found in linear time. 
\end{IEEEproof}

\medskip

As said before, Lemma \ref{lemma:pezzettini} is  an important  technical  tool of our main algorithm. Therefore,
in \textbf{Algorithm \ref{algo:Lemma}} we give an efficient way to compute the value $z^{(d)}$ and the set 
of indices $I$.

\begin{algorithm}[ht!]
\small 
{\sc  Min-Entropy-Joint-Distr}($\p , \q$)\\ 
{\bf Input:} prob.\ distributions $\p = (p_1, \dots, p_n)$ and $\q= (q_1, \dots, q_n)$\\
{\bf Output:} An $n \times n$ matrix $\mathbf{M} = [m_{i\,j}]$ s.t.\ $\sum_{j} m_{i\,j} = p_i$ and $\sum_{i} m_{i\,j} = q_j.$
\begin{algorithmic}[1]
\STATE{{\bf for} $i=1,\dots, n$ and $j=1, \dots, n$  {\bf set} $m_{i \, j} \leftarrow 0$} \label{newalgo:initiate-first}
\STATE{{\bf if} $\p \neq \q$, let $i = \max \{j \mid p_j \neq q_j\}$; 
{\bf if} $p_i < q_i$ {\bf swap} $\p \leftrightarrow \q$} \label{newalgo:initiate-swap}
\STATE{$\z = (z_1, \dots, z_n) \leftarrow \p \land \q$}
\STATE{{\bf for} $i=1,\dots, n$  {\bf set} $m_{i\, i} \leftarrow z_i$} \label{newalgo:initiate-last}
\STATE{$i \leftarrow n$}
\WHILE{$i \geq 1$} \label{newalgo:while-start}
\IF{$\sum_{k=i}^n m_{k\, i} > q_i$}  \label{newalgo:if1-start}
\STATE{($z_{i}^{(d)},  z_{i}^{(r)}, I$) $\leftarrow$  \label{newalgo:if1-1}
{\sc  Lemma\ref{lemma:pezzettini}}($z_i, q_i, [m_{1 \, i}, m_{2\, i}, \dots, m_{n\, i}]$)}   
\STATE{$m_{i\, i} \leftarrow z^{(d)}_i, \, m_{i, i-1}  \leftarrow z^{(r)}_i$} \label{split-1}
\FOR{{\bf each} $k \not \in I \cup\{i\}$}  
\STATE{$m_{k\, i-1} \leftarrow m_{k\, i}$} \label{move-1-1}
\STATE{$m_{k\, i} \leftarrow 0$}   \label{move-1-2}
\ENDFOR
\ENDIF  \label{newalgo:if1-end}
\IF{$\sum_{k=i}^n m_{i\, k} > p_i$}   \label{newalgo:if2-start}
\STATE{($z_{i}^{(d)},  z_{i}^{(r)}, I$) $\leftarrow$  
{\sc  Lemma\ref{lemma:pezzettini}}($z_i, p_i, [m_{i \, 1}, m_{i\, 2}, \dots, m_{i\, n}]$)}   
\STATE{$m_{i\, i} \leftarrow z^{(d)}_i, \, m_{i-1, i}  \leftarrow z^{(r)}_i$}  \label{split-2}
\FOR{{\bf each} $k \not \in I \cup\{i\}$}  
\STATE{$m_{i-1\, k} \leftarrow m_{i\, k}$} \label{move-2-1}
\STATE{$m_{i\, k} \leftarrow 0$} \label{move-2-2}
\ENDFOR
\ENDIF   \label{newalgo:if2-end}
\STATE{$i \leftarrow i-1$}
\ENDWHILE
\end{algorithmic}
\caption{The Min Entropy Joint Distribution Algorithm}
\label{algo:Algonewnew}
\end{algorithm}

\begin{algorithm}[ht!]
\small
{\sc  Lemma\ref{lemma:pezzettini}}($z, x, A[i\dots j]$)\\ 
{\bf Input:} reals $z > 0, \, x \geq 0,$ {  and $A[i \dots j] \geq 0$ 
s.t.  $\sum_{k} A[k] + x \geq z$ } \\
{\bf Output:} $z^{(d)}, z^{(r)} \geq 0,$ and $I \subseteq \{i, i+1, \dots, j\}$ s.t. $z^{(d)} + z^{(r)} = z,$ and $z^{(d)} + 
\sum_{\ell \in I} A[\ell] = x.$

\begin{algorithmic}[1]
\STATE{$k \leftarrow i, \, I \leftarrow \emptyset, \, sum \leftarrow 0$}  
\WHILE{$k \leq j$ {\bf and} $sum + A[k] < x$}
\STATE{$I \leftarrow I \cup \{k\}, \, sum \leftarrow sum + A[k], \, k \leftarrow k+1$} 
\ENDWHILE
\STATE{$z^{(d)} \leftarrow x - sum, \, z^{(r)} \leftarrow z - z^{(d)}$}
\STATE{{\bf return} ($z^{(d)}, z^{(r)}, I$)}
\end{algorithmic}
\caption{The procedure implementing Lemma \ref{lemma:pezzettini}}
\label{algo:Lemma}
\end{algorithm}

By padding the probability distributions with the appropriate
number of $0$'s, we can assume that both  $\p,\q\in \PP_n$. We are now ready to present our main algorithm.
 The pseudocode is given in \textbf{Algorithm 1}. 
Since the description of \textbf{Algorithm 1} might look complicated, we
see fit to illustrate and comment its behavior with a worked out example.
The reader is advised to go through the content of Section  \ref{informal} before
reading the formal proofs of Section \ref{sec:proof}. 

\subsection{{How  \textbf{Algorithm 1} works: An informal   description of its functioning and a numerical example}}\label{informal}

At any point during the execution of the algorithm, 
we say that $\q$ is $i$-satisfied if  the 
sum of the entries on columns $i, i+1, \dots, n$ 
of the matrix the algorithm is constructing,
 is equal  to $q_i + q_{i+1}+ \dots +  q_{n}$ 
Analogously, we say that $\p$ is 
$i$-satisfied if  the 
sum of the entries on rows 
$i, i+1, \dots, n$ is equal  
to $p_i + p_{i+1}+ \dots +  p_{n}$. 
Clearly, a matrix  $\mathbf{M} \in \CC(\p, \q)$  if and only if
it holds that  both 
$\p$ and $\q$ are $i$-satisfied for each $i = 1, \dots, n$.

Let $\z$ be the vector defined in \textbf{Fact \ref{glb}}, and $\mathbf{M}_{\z}$ be a matrix defined by 
setting $\mathbf{M}_{\z}[i,i] = z_i$ and setting all the other entries to zero. 
The basic observation is that for the matrix  $\mathbf{M}_{\z}$
 either  $\p$ or $\q$ 
is $i$-satisfied, for each $i=1, \dots, n$, (but not necessarily  both). In addition, 
every constraint which is not satisfied, coincides with an overflow, i.e., 
if for instance for some $i$ we have that for $\mathbf{M}_{\z}$ defined above 
$\p$ is not $i$-satisfied, it is
necessarily the case that the sum of rows $i, i+1, \dots, n$ of $\mathbf{M}_{\z}$  is strictly greater than 
$p_i + p_{i+1} + \dots + p_n.$ 

We can understand our algorithm as working on how to modify $\mathbf{M}_{\z}$ in order to achieve $i$-satisfiability
for both $\p$ and $\q$ for each $i =1, \dots, n,$ by splitting \emph{in at most two parts} each diagonal element. 
The algorithm processes the vector $\z$ from the smallest component $z_n$ to the largest $z_1$. 
For $i=n, \dots, 1$ it keeps $z_i$ in the diagonal entry $\mathbf{M}[i,i]$ as long as
both $\p$ and $\q$ are $i$-satisfied.
 
When, e.g., $\q$ is not $i$-satisfied, it must be necessarily overflowed, i.e., the sum of the
components on the $i$-th column is larger than $q_i.$ 
Then, the algorithm's action is equivalent to  
removing the surplus from the $i$-th column  
and place it onto the column $i-1$  
so that $\q$ becomes $i$-satisfied and  $\p$ remains $i$-satisfied, as the mass moved is still on the same rows.  

This operation can be accomplished using Lemma 3, i.e., by selecting  
a subset of the non-zero components on column $i$ together with $0 < z_i' < z_i$
so that their sum is equal to $q_i$. Keep this mass on column $i$ and move the 
remaining components and the left over of $z_i$ to column $i-1.$ In this process
only $z_i$ gets split.

Analogously, when $\p$ is not $i$-satisfied, it must be necessarily overflowed, i.e., the sum of the
components on the $i$-th row is larger than $p_i.$ 
Then, the algorithm's action is equivalent to  
removing the surplus from the $i$-th row  
and place it onto the row $i-1$  
so that $\p$ becomes $i$-satisfied and  $\q$ remains $i$-satisfied, as the mass moved 
is still on the same columns.  

This operation is again accomplished using Lemma 3: select 
a subset of the non-zero components on row $i$ together with $0 < z_i' < z_i$
so that their sum is equal to $p_i$. Keep this mass on row $i$ and move the 
remaining components and the left over of $z_i$ to row $i-1.$ Again 
in this process only $z_i$ gets split.

\bigskip
Let us consider the following example:  
Let $n = 6$ and $\p = (0.4, 0.3, 0.15, 0.08, 0.04, 0.03)$ and 
$\q = (0.44, 0.18, 0.18, 0.15, 0.03, 0.02),$ be the two probability 
distributions for which we are seeking a coupling of minimum entropy.
We have
$\z = \p \land \q = (0.4, 0.22, 0.18, 0.13, 0.04, 0.03).$

In the first iteration, we process the entry $(6,6)$ containing $z_6$ (indicated in bold, below). In the matrix $\mathbf{M}_z$ (below)
we have that $\p$ is $6$-satisfied but $q_6$ is overflowed. Therefore, we split $z_6$ into $0.2 = q_6$ and $0.1$ and
leave the former as entry  $m_{6\,6}$ and the make the latter be entry 
$m_{6\,5},$ obtaining the matrix $\mathbf{M}^{(6)}$ on the right. 
The underlined values represent the mass that has been moved from one column to the next one on the left.
$${  \mathbf{\scalebox{0.8}{M}_z}} = \left (\begin{matrix}
 0.4 &  &  &  &  & 0\\
 & 0.22 &  &   &  & 0 \\
 &  &  0.18 &  &  & 0\\
 &  &  & 0.13 &  & 0 \\
 &  &  &  & 0.04 & 0 \\
0 & 0 & 0 & 0 & 0 & {\bf 0.03} 
\end{matrix}\right )$$

$$
{  \mathbf{\scalebox{0.8}{M}^{(6)}}} = \left (\begin{matrix}
 0.4 &  &  &  &  & 0\\
 & 0.22 &  &   &  & 0 \\
 &  &  0.18 &  &  & 0\\
 &  &  & 0.13 &  & 0 \\
 &  &  &  & 0.04 & 0 \\
0 & 0 & 0 & 0 & \underline{0.01} & 0.02 \end{matrix}\right )
$$

Then, we process entry $(5,5)$ containing $z_5$ (indicated in bold, below). In $\mathbf{M}^{(6)}$ we now have that 
 $\p$ is $5$-satisfied but $q_5$ is overflowed. Therefore, we apply Lemma 3 to column 5, in order to find 
a split of $z_5$ and some of the other components of column 5 whose total sum is equal to $q_5$ and we 
move the remaining mass to column 4. Splitting $z_5$ into $0.2 + 0.2$ we obtain the matrix $\mathbf{M}^{(5)}$ 
on the right. 
The underlined values represent 
the mass that has been moved from one column to the next one on the left.

$${  \mathbf{\scalebox{0.8}{M}^{(6)}}} = \left (\begin{matrix}
 0.4 &  &  &  &  & 0\\
 & 0.22 &  &   &  & 0 \\
 &  &  0.18 &  &  & 0\\
 &  &  & 0.13 &  & 0 \\
 &  &  &  & {\bf 0.04}  &  0\\
0 & 0 & 0 & 0 & \underline{0.01} & 0.02 
\end{matrix}\right )$$

$$
{  \mathbf{\scalebox{0.8}{M}^{(5)}}} = \left (\begin{matrix}
0.4 &  &  &  & 0 & 0\\
 & 0.22 &  &   & 0 & 0 \\
 &  &  0.18 &  & 0 & 0\\
 &  &  & 0.13 & 0 & 0 \\
0 & 0 & 0 & \underline{0.02} & 0.02  &  0\\
0 & 0 & 0 & 0 & 0.01 & 0.02 
\end{matrix}\right )
$$

Then, we process entry $(4,4)$ containing $z_4$ (indicated in bold, below). In $\mathbf{M}^{(5)}$ we now have that
$\q$ is $4$-satisfied but $p_4$ is overflowed. Therefore, we apply Lemma 3 to row 4, in order to find 
a split of $z_4$ such that one part is equal to $p_4$ and we 
move the remaining mass to row 3. Splitting $z_4$ into $0.8 + 0.5$ we obtain the matrix $\mathbf{M}^{(4)}$ on the right. 
The
underlined values represent the mass that has been moved.

$${  \mathbf{\scalebox{0.8}{M}^{(5)}}} = \left (\begin{matrix}
  0.4 &  &  &  & 0 & 0\\
 & 0.22 &  &   & 0 & 0 \\
 &  &  0.18 &  & 0 & 0\\
&  &  & {\bf 0.13} & 0 & 0\\
0 & 0 & 0 & \underline{0.02} & 0.02  &  0\\
0 & 0 & 0 & 0 & 0.01 & 0.02 
\end{matrix}\right )$$

$$
{  \mathbf{\scalebox{0.8}{M}^{(4)}}} = \left (\begin{matrix}
  0.4 &  &  & 0 & 0 & 0\\
 & 0.22 &  &  0 & 0 & 0 \\
 &  & 0.18 & \underline{0.05} & 0 & 0\\
0 & 0 & 0 & 0.08 & 0 & 0\\
0 & 0 & 0 & 0.02 & 0.02  &  0\\
0 & 0 & 0 & 0 & 0.01 & 0.02 
\end{matrix}\right )
$$

Then, we process entry $(3,3)$ containing  $z_3$ (indicated in bold, below). In $\mathbf{M}^{(4)}$ we now have that
  $\q$ is $3$-satisfied but $p_3$ is overflowed. Therefore, we apply Lemma 3 to row 4, in order to find 
a split of $z_3$ and some of the other components of row 3 whose total sum is equal to $p_3$ and we 
move the remaining mass to row 2.
If we split $z_3$ into $0.15 + 0.03$ we obtain the matrix $\mathbf{M}^{(3)}$ on the right.
The
underlined values represent the mass that has been moved from one row to the next one above.

$${  \mathbf{\scalebox{0.8}{M}^{(4)}}} = \left (\begin{matrix}
  0.4 &  &  & 0 & 0 & 0\\
 & 0.22 &  &  0 & 0 & 0 \\
 &  & {\bf 0.18} & \underline{0.05} & 0 & 0\\
0 & 0 & 0 & 0.08 & 0 & 0\\
0 & 0 & 0 & 0.02 & 0.02  &  0\\
0 & 0 & 0 & 0 & 0.01 & 0.02 
\end{matrix}\right )$$

$$
{  \mathbf{\scalebox{0.8}{M}^{(3)}}} = \left (\begin{matrix}
 0.4 &  & 0 & 0 & 0 & 0\\
 & 0.22 & \underline{0.03} & \underline{0.05} & 0 &  0\\
0 & 0 & 0.15 & 0 & 0 & 0\\
0 & 0 & 0 & 0.08 & 0 & 0\\
0 & 0 & 0 & 0.02 & 0.02  &  0\\
0 & 0 & 0 & 0 & 0.01 & 0.02 
\end{matrix}\right )
$$

Then, we process entry $(2,2)$ containing $z_2$ (indicated in bold, below). In $\mathbf{M}^{(3)}$ we now have that
$\p$ is $2$-satisfied but $q_2$ is overflowed. Therefore, we apply Lemma 3 to column 2, in order to find 
a split of $z_2$ and some of the other components of column 2 whose total sum is equal to $q_2$ and we 
move the remaining mass to column 1.
If we split $z_2$ into $0.18 + 0.04$ we obtain the matrix $\mathbf{M}^{(2)}$ on the right.
The
underlined values represent the mass that has been moved from one column to the next one on the left.

{$${ \mathbf{\scalebox{0.8}{M}^{(3)}}} = \left (\begin{matrix}
0.4 &  & 0  & 0 & 0 & 0\\
 & {\bf 0.22}  & \underline{0.03} & \underline{0.05} & 0 &  0\\
0 & 0 & 0.15 & 0 & 0 & 0\\
0 & 0 & 0 & 0.08 & 0 & 0\\
0 & 0 & 0 & 0.02 & 0.02  &  0\\
0 & 0 & 0 & 0 & 0.01 & 0.02 
\end{matrix}\right )$$

$$
{  \mathbf{\scalebox{0.8}{M}^{(2)}}} = 
\quad \left (\begin{matrix}
0.4 & 0 & 0 & 0 & 0 & 0\\
\underline{0.04} & 0.18  & 0.03 & 0.05 & 0 &  0\\
0 & 0 & 0.15 & 0 & 0 & 0\\
0 & 0 & 0 & 0.08 & 0 & 0\\
0 & 0 & 0 & 0.02 & 0.02  &  0\\
0 & 0 & 0 & 0 & 0.01 & 0.02 
\end{matrix}\right )
$$
}

Finally, we process entry $(1,1)$ containing  $z_1$ (indicated in bold, below). In $\mathbf{M}^{(2)}$ we have that
both  $\p$ and $\q$ are $1$-satisfied. Therefore we get the unmodified matrix 
$\mathbf{M}^{(1)}$ on the right which is our joint distribution. Notice that each component of $\z$ 
has been split at most into two parts. In particular only when $z_i$ is processed the first time it might 
get split, while the other components (obtained by 
the previous subdivision of some other components of $\z$)
 might be relocated but not chunked again. 

$${ \mathbf{\scalebox{0.8}{M}^{(2)}}} =
 \left (\begin{matrix}
{\bf 0.4} & 0 & 0 & 0 & 0 & 0\\
\underline{0.04} & 0.18  & 0.03 & 0.05 & 0 &  0\\
0 & 0 & 0.15 & 0 & 0 & 0\\
0 & 0 & 0 & 0.08 & 0 & 0\\
0 & 0 & 0 & 0.02 & 0.02  &  0\\
0 & 0 & 0 & 0 & 0.01 & 0.02 
\end{matrix}\right )$$

$$
{  \mathbf{\scalebox{0.8}{M}^{(1)}}} =
\quad  \left (\begin{matrix}
0.4 & 0 & 0 & 0 & 0 & 0\\
0.04 & 0.18  & 0.03 & 0.05 & 0 &  0\\
0 & 0 & 0.15 & 0 & 0 & 0\\
0 & 0 & 0 & 0.08 & 0 & 0\\
0 & 0 & 0 & 0.02 & 0.02  &  0\\
0 & 0 & 0 & 0 & 0.01 & 0.02 
\end{matrix}\right ) = 
{  \mathbf{\scalebox{0.8}{M}}} 
$$

\medskip 

\subsection{The proof of correctness of  \textbf{Algorithm \ref{algo:Algonewnew}}}\label{sec:proof}

The following theorem shows the correctness of \textbf{Algorithm \ref{algo:Algonewnew}}. In particular,
the equalities in (\ref{main:prop3}), for the case $i=1$, imply that the matrix built by the algorithm is a coupling of $\p$ and $\q$.

\begin{theorem}
For each $i = n, n-1, \dots, 1$ at the beginning of iteration $i$ of the main {\bf while} loop the following conditions hold
\begin{enumerate}
\item \label{main:prop1} for each $i' \leq i$ we have $$\sum_{\ell = i'}^ n  \sum_{k=1}^n m_{\ell\, k} =  
\sum_{\ell = i'}^ n  \sum_{k=1}^n m_{k\, \ell} =  \sum_{\ell = i'}^n z_{\ell},$$
\item  \label{main:prop2} exactly one of the following holds
\begin{enumerate} 
\item \label{main:prop2-1} $\sum_{k=1}^n m_{k\, i} = q_i$ and  $\sum_{k=1}^n m_{i\, ki} = p_i$
\item \label{main:prop2-2} $\sum_{k=1}^n m_{k\, i} > q_i$ and  $\sum_{k=1}^n m_{i\, ki} = p_i$
\item \label{main:prop2-3} $\sum_{k=1}^n m_{k\, i} = q_i$ and  $\sum_{k=1}^n m_{i\, ki} > p_i$
hence, at most one of the {\bf if} conditions is true.
\end{enumerate}
\end{enumerate}
Moreover at the end of the iteration $i,$  
for each $i' \geq i$ it holds that
\begin{equation} \label{main:prop3}
\sum_{k=1}^n m_{k \, i'} = q_{i'} \qquad \mbox{and} \qquad  \sum_{k=1}^n m_{i' \, k} = p_{i'}.
\end{equation}
\end{theorem}
\begin{IEEEproof}
We prove the statement by reverse induction. For $i = n$, due to the initialization in lines 
\ref{newalgo:initiate-first} and \ref{newalgo:initiate-last} we have that for each $i' \leq n$ the only 
non-zero entry in row $i'$ and in column $i'$ is $m_{i'\, i'} = z_{i'}$ and \ref{main:prop1} holds.

By definition we have $m_{n\,n} = z_{n} = \max\{p_n, q_n\} = p_n,$ since,  
by the initialisation in line \ref{newalgo:initiate-swap} we can assume $p_n \geq q_n$.
Therefore, either
$m_{n\, n} = p_n = q_n$ and \ref{main:prop2-1}) holds; or 
$m_{n\, n} = p_n > q_n$ and \ref{main:prop2-2}) holds. 
Thus, \ref{main:prop2}) holds.

Finally, if \ref{main:prop2-1}) holds during the iteration $n$ then no modification of the 
matrix entries is performed and at the end of the iteration equation (\ref{main:prop3}) holds.
Otherwise, as already observed, because of the initialization in line \ref{newalgo:initiate-swap},
 we have $z_n = p_n > q_n$. Then, as a result of the call to procedure {\sc Lemma3}($z_n, q_n, A = \emptyset$) 
 we will  have $z^{(d)}_n = q_n,\, z^{(r)}_n = p_n - q_n$ and the modifications to the matrix entries
 $m_{n,n} = z^{(d)}_n = q_n$ and $m_{n,n-1} = z^{(r)}_n = p_n - q_n,$ from which (\ref{main:prop3}) 
 holds at the end of the iteration as desired. This settles the induction base. 
 
Let us now assume that the claims hold for iteration $i+1$ and prove it for iteration $i$.

\medskip

\noindent
\ref{main:prop1}. By induction hypothesis, at the beginning of iteration $i+1$
 for each $i' \leq i+1$, hence in particular for each $i' \leq i$ it holds that
\begin{equation} \label{previous:1}
\sum_{\ell = i'} \sum_{k=1}^n m_{\ell\, k} = \sum_{\ell = i'} \sum_{k=1}^n m_{k \, \ell} = \sum_{\ell = i'} z_{\ell}. 
\end{equation}
During iteration $i+1$ the only possible changes to entries of the matrix are either in rows $i$ and $i+1$ 
(when the {\bf if} at line \ref{newalgo:if1-start} is satisfied) or 
in columns $i$ and $i+1$ (when the {\bf if} at line \ref{newalgo:if2-start} is satisfied). 
Moreover, such modifications do not change the total probability mass in 
rows $i$ and $i+1$ and the total probability mass in column $i$ and $i+1$, i.e., the sums
$\sum_{k=1}^n (m_{k\, i} + m_{k\, i+1})$  and $\sum_{k=1}^n (m_{i\, k} + m_{i+1\, k})$ remain 
unchanged during iteration $i+1$. It follows that at the beginning of iteration $i$ equality (\ref{previous:1}) 
still holds for each $i' \leq i.$ This settles the inductive steps for property \ref{main:prop1}.

\medskip
\noindent
\ref{main:prop2}.
By induction hypothesis from \ref{main:prop1} with $i' = i < i+1$ we have
\begin{equation} \label{ind-2:1}
\sum_{\ell = i}^n \sum_{k=1}^n m_{\ell\, k} = \sum_{\ell = i}^n \sum_{k=1}^n m_{k\, \ell} = \sum_{\ell = i}^n z_{\ell}.
\end{equation}
By induction hypothesis, we also have that for each $\ell = i+1, \dots, n,$
\begin{equation} \label{ind-2:2}
\sum_{k=1}^n m_{\ell\, k} = p_{\ell} \quad \mbox{and} \quad \sum_{k=1}^n m_{k\, \ell} = q_{\ell} 
\end{equation}
From equations (\ref{ind-2:1})-(\ref{ind-2:2}) together with (\ref{z-suffix-sum}) we get 
\begin{equation} \label{cons2:1}
\max\{\sum_{\ell = i}^n p_{\ell}, \sum_{\ell = i}^n q_{\ell}\} = \sum_{\ell = i}^n z_{\ell} = 
\sum_{\ell = i+1}^n p_{\ell} + \sum_{k = 1}^n m_{i\, \ell}
\end{equation}
and
\begin{equation} \label{cons2:2}
\max\{\sum_{\ell = i}^n p_{\ell}, \sum_{\ell = i}^n q_{\ell}\} = \sum_{\ell = i}^n z_{\ell} = 
\sum_{\ell = i+1}^n q_{\ell} + \sum_{k = 1}^n m_{\ell \, i}.
\end{equation}
Therefore, (a) if $\sum_{\ell = i}^n z_{\ell}  = \sum_{\ell = i}^n q_{\ell} =  \sum_{\ell = i}^n p_{\ell}$
then from (\ref{cons2:1}) we have $p_i = \sum_{k=1}^n m_{i\,k}$ and 
from (\ref{cons2:2}) we have $q_i = \sum_{k=1}^n m_{k\, i}.$

(b) If $\sum_{\ell = i}^n z_{\ell}  = \sum_{\ell = i}^n p_{\ell} >  \sum_{\ell = i}^n q_{\ell}$
then from (\ref{cons2:1}) we have $p_i = \sum_{k=1}^n m_{i\,k}$ and 
from (\ref{cons2:2}) we have $q_i < \sum_{k=1}^n m_{k\, i}.$

(c) If $\sum_{\ell = i}^n z_{\ell}  = \sum_{\ell = i}^n q_{\ell} >  \sum_{\ell = i}^n p_{\ell}$
then from (\ref{cons2:1}) we have $p_i < \sum_{k=1}^n m_{i\,k}$ and 
from (\ref{cons2:2}) we have $q_i = \sum_{k=1}^n m_{k\, i}.$

Exactly one of these three cases is possible, which proves the induction step for 
\ref{main:prop2}).

\medskip

\noindent 
Let us now prove the induction step for (\ref{main:prop3}). 

Assume first that during the iteration $i$ case \ref{main:prop2-1}) applies. From the previous
point, this means that $\sum_{\ell = i}^n z_{\ell} = \sum_{\ell=i}^n q_{\ell} = \sum_{\ell=i}^n p_{\ell}$. Then, 
none of the two {\bf if} compounds (lines \ref{newalgo:if1-start}-\ref{newalgo:if1-end} and 
lines \ref{newalgo:if2-start}-\ref{newalgo:if2-end}) are executed and no matrix entry is changed in this iteration.
As a result at the end of the iteration we have that, for each 
$i' > i$ the formula (\ref{main:prop3}) holds by induction hypothesis.  Moreover it also holds for $i' = i$ since, from \ref{main:prop1})
and \ref{main:prop2-1}) we have
\begin{eqnarray*}
\sum_{k=1}^n m_{k\, i} &=& \sum_{\ell = i}^n z_{\ell} - \sum_{\ell=i+1}^n \sum_{k=1}^n m_{k\, \ell} \\
&=& \sum_{\ell = i}^n q_{\ell} - \sum_{\ell=i+1}^n q_{\ell} = q_i
\end{eqnarray*}
and analogously
\begin{eqnarray*}
\sum_{k=1}^n m_{i\, k} &=& \sum_{\ell = i}^n z_{\ell} - \sum_{\ell=i+1}^n \sum_{k=1}^n m_{\ell\, k} \\
&=& \sum_{\ell = i}^n p_{\ell} - \sum_{\ell=i+1}^n p_{\ell} = p_i
\end{eqnarray*}

Assume now that during the iteration $i$ case \ref{main:prop2-2}) applies (the case \ref{main:prop2-3}) can be dealt with 
symmetrically). Then, the {\bf if} compound in lines \ref{newalgo:if1-start}-\ref{newalgo:if1-end} is executed. As a 
result, values $z^{(d)}_i, z^{(r)}_i$ and set $I \subseteq [n]$ are computed such that $z^{(d)}_i + z^{(r)}_i = z_i$ and 
$z^{(d)}_i + \sum_{k \in I} m_{k\, i} = q_i$. 
Before the assignments in line 9 and the execution of the {\bf for} loop, we had that 
\begin{align*}
\sum_{k=1}^n m_{k\,i} &= m_{i\,i} + \sum_{k\neq i} m_{k\,i} = z_i + \sum_{k \in I} m_{k\,i} + \sum_{k\not \in I\cup\{i\}} m_{k\,i}\\
                      &= z_i^{(d)} + \sum_{k \in I} m_{k\,i} + z^{(r)}_i + \sum_{k\not \in I\cup\{i\}} m_{k\,i}.
\end{align*}
After the assignments in line 9 and the execution of the {\bf for} loop, the mass in the  last two terms is moved to 
column $i-1$, but without changing the row. Therefore the row sums do not change and the column sums of 
column $i$ and $i-1$ change so that $\sum_{k=1}^n m_{k\, i} = z^{(d)}_i + \sum_{k \in I} m_{k\, i} = q_i$ as desired. 

Finally, it is possible that during the iteration $i$ case \ref{main:prop2-3}) applies. The analysis for this case is 
analogous to the one for the previous case, from which it can be easily obtained by symmetry swapping the roles of 
rows and columns, taking into account that we have to consider the result of the operations executed within 
the {\bf if} compound in lines \ref{newalgo:if2-start}-\ref{newalgo:if2-end}.

The proof is complete.
\end{IEEEproof}

\subsection{The { guaranteed additive gap} of  \textbf{Algorithm \ref{algo:Algonewnew}}}\label{sec:proof-old}

We are now ready to formally prove our main result.

\begin{theorem}\label{th:main}
For any $\p, \q \in \PP_n$, Algorithm \ref{algo:Algonewnew} outputs {\em in 
polynomial time} an $\mathbf{M}\in \CC(\p, \q)$ such that 
\begin{equation} \label{eq:maintheo}
H(\mathbf{M})\leq H(\p\land \q)+1. 
\end{equation}
\end{theorem}
\begin{IEEEproof}
It is not hard to see that the  values of all non-zero entry of the matrix  $\mathbf{M}$ are initially set 
in line \ref{newalgo:initiate-last} and then in lines  \ref{split-1} and  \ref{split-2}---in fact, 
the assignments  
in lines \ref{move-1-1}-\ref{move-1-2} and \ref{move-2-1}-\ref{move-2-2} have the effect of shifting 
by one column to the left or by one row up values 
that had been fixed at some point earlier in lines \ref{split-1} and  \ref{split-2}.
Therefore, all the final non-zero entries of $\mathbf{M}$ can be partitioned into $n$ pairs $z_j^{(r)}, z_j^{(d)}$ with 
$z_j^{(r)} + z_j^{(d)} = z_j$ for $j = 1, \dots, n$. By using the standard assumption $0 \log \frac{1}{0} = 0$ and 
applying Jensen inequality we have
\begin{eqnarray*}
H(\mathbf{M}) &=& 
\sum_{j=1}^n z_j^{(r)} \log \frac{1}{z_j^{(r)}} + z_j^{(d)} \log \frac{1}{z_j^{(d)}} \\
&=&\sum_{j=1}^n \left [ z_j\left (\frac{z_j^{(r)}}{z_j}\log \frac{1}{z_j^{(r)}}+ \frac{z_j^{(d)}}{z_j}\log \frac{1}{z_j^{(d)}} \right )\right ]    \\
&\leq& {\sum_{j=1}^n {z_j} \log \frac{2}{z_j}} = H(\z) + 1
\end{eqnarray*}
which concludes the proof of the { bound on the additive gap} guarantee{d by} Algorithm \ref{algo:Algonewnew}.
Moreover, one can see that Algorithm \ref{algo:Algonewnew} can be implemented so to run in $O(n^2)$ time.
For the time complexity of the algorithm we observe the following easily verifiable fact:
\begin{itemize}
\item the initialization in line \ref{newalgo:initiate-first} takes $O(n^2)$;
\item the condition in line \ref{newalgo:initiate-swap} can be easily verified in 
$O(n)$ which is also the complexity of swapping $\p$ with $\q$, if needed;
\item the vector $z = \p \land \q$ can be computed in $O(n)$, once the suffix sums 
$\sum_{j=k}^n p_j, \, \sum_{j=k}^n q_j$ ($k = 1, \dots, n$) have been precomputed (also doable in $O(n)$); 
\item the main {\bf while} loop is executed $n$ times and all the operations
executed in an iteration are easily upper bounded by $O(n)$. The most expensive are 
 the  calls to the procedure {\bf Lemma3}, and the {\bf for}-loops. All these take $O(n)$.
Therefore, the overall running time of the {\bf while} loop is also $O(n^2)$.
\end{itemize}
Therefore we can conclude that the time complexity of Algorithm \ref{algo:Algonewnew} is $O(n^2),$ hence 
polynomial time. 
\end{IEEEproof}

\subsection{Improving the time complexity} \label{section:improvedtime}
We note that the time complexity of Algorithm \ref{algo:Algonewnew} can be improved if 
we build the coupling ${\bf M}$ in sparse form, i.e., as the set of values $\{({\bf M}[i,j], (i,j)) \mid {\bf M}[i,j] \neq 0\}$ 
containing only the non-zero entr{ies} of ${\bf M}$ together with their coordinates.

We can keep the moved masses, i.e., the pieces $z_i^{(r)}$ that are iteratively moved from one column to the previous one, in line \ref{move-1-1} 
(respectively,  from one row to the previous one, in line \ref{move-2-1})
in a priority queue ${\cal Q}$. For each such element we store in the priority queue its row index (resp. column index) and its mass. 
With a standard implementation of a priority queue, we can   then 
efficiently find the value of the minimum mass stored in constant time $O(1)$ (we refer to this operation as {\sc Min(${\cal Q}$})
and extract the minimum mass in time logarithmic in the number of elements stored in the priority queue
(we refer to this operation as {\sc ExtractMin(${\cal Q}$}))\cite{Cormen}.
Accordingly, procedure Lemma 3 amounts to iteratively extract from the priority queue the smallest mass as long as the queue is not empty and the sum of the masses
extracted do not overcomes $\min\{p_i, q_i\}$ . Whenever we split $z_i$, we insert $z_i^{(r)}$  into the priority queue 
(this operation can also be implemented to require time logarithmic in the size of the queue; we refer to it as {\sc Insert(${\cal Q},(z_i^{(r)}, i)$)}). 

At any time, the priority queue will contain 
$O(n)$ elements. Therefore, each insertion ({\sc Insert}) and extraction ({\sc ExtractMin}) from the priority queue takes $O(\log n)$ time. Moreover, since each 
element enters the queue at most once, the overall time of {\em all} insertion and extraction operations is upper bounded by $O(n \log n)$. 
The remaining 
part of the algorithm takes $O(n)$, apart from the possible initial sorting of the two distribution, adding another $O(n \log n)$ term. 
Therefore, the resulting implementation has complexity $O(n \log n)$. 

We report in appendix a pseudocode of such implementation, where, 
for the sake of a clearer description, we
use two priority queues, ${\cal Q}^{(row)}, {\cal Q}^{(col)},$ storing masses moved among rows and masses moved among columns respectively.

\section{Extending the Results  to other Entropy Measures}
Our approach to prove entropic inequalities via majorization theory seems quite 
powerful. Indeed, it  allows us to extend our  results to different kind of entropies, 
with no  additional effort. As an example, let us consider the order $\alpha$ R\'enyi entropy \cite{renyi2} 
of a probability distribution $\p=(p_1, \ldots , p_n)$, defined as
\begin{equation}\label{eq:re0}
 H_\alpha(\p)=\frac{1}{1-\alpha}\log \sum_{i=1}^np_i^\alpha,
\end{equation}
where $\alpha\in (0,1)\cup (\,\infty)$. It is well known that the R\'enyi entropy is Schur-concave, for all the values of the
parameter $\alpha$ \cite{HV2015}. Therefore, we immediately have the
analogous of Lemma \ref{lemma:HM>Hpandq} of Section \ref{sec:pre}. 

\begin{lemma}\label{lemma:HM>HpandqR}
\emph{For any} $\p$ and $\q$,  for any $\mathbf{M}\in \CC(\p, \q)$ and $\alpha\in (0,1)\cup (\,\infty)$, it holds that 
\begin{equation}\label{eq:HM>HpandqR}
H_\alpha(\mathbf{M})\geq H_\alpha(\p\land \q).
\end{equation}
\end{lemma} 

We now prove the analogous of Theorem \ref{th:main}, that is
\begin{theorem}\label{th:mainR}
For any $\p, \q \in \PP_n$, Algorithm \ref{algo:Algonewnew} outputs {\em in 
polynomial time} an $\mathbf{M}\in \CC(\p, \q)$ such that 
\begin{equation} \label{eq:maintheoR}
H_\alpha(\mathbf{M})\leq H_\alpha(\p\land \q)+1\leq \min_{\mathbf{N}\in \CC(\p, \q)} H_\alpha(\mathbf{N})+1.
\end{equation}
\end{theorem}
\begin{IEEEproof}
Let $\mathbf{M}$ be the matrix constructed by our Algorithm \ref{algo:Algonewnew}, and let $\alpha\in (0,1)$.
Proceedings as in Theorem \ref{th:main} (and with the same notations), we have:
\begin{align*}
H_\alpha(\mathbf{M}) &= \frac{1}{1-\alpha}\log \sum_{j=1}^n \left [\left (z_j^{(r)} \right )^\alpha  + \left (z_j^{(d)}\right )^\alpha \right ] \\
                     &= \frac{1}{1-\alpha}\log \sum_{j=1}^n 2 \left [\frac{1}{2}\left (z_j^{(r)} \right )^\alpha  + \frac{1}{2}\left (z_j^{(d)}\right )^\alpha \right ]\\
                     &\leq \frac{1}{1-\alpha}\log \sum_{j=1}^n 2\left (\frac{1}{2}z_j^{(r)}+\frac{1}{2}z_j^{(d)} \right )^\alpha \\
										     & \qquad\quad\qquad \mbox{(by the Jensen inequality applied to $x^\alpha$)}\\
									   &= \frac{1}{1-\alpha}\log \sum_{j=1}^n 2 \left (\frac{z_j}{2}\right )^\alpha 
										               \quad   \mbox{(since  $z_j^{(r)}+z_j^{(d)}=z_j$)}\\
										 &= H_\alpha(\z)+1=H_\alpha(\p\land \q)+1.
\end{align*}
The proof for the case $\alpha\in (1,\infty)$ is the same, by noticing that for $\alpha>1$ the Jensen inequality goes into the opposite
direction and that $1/(1-\alpha)<0$.
\end{IEEEproof}

\bigskip
One can extend our results also to other entropies, like the Tsallis entropy \cite{tsallis}, using its  Schur-concavity 
property proved in \cite{fu}. The mathematical details can be easily worked out by the motivated reader.

\section{An Extension  to Multivariate Distributions}\label{sec:more}

In this section we will show how the algorithm {\sc  Min-Entropy-Joint-Distr} can be used to 
{ guarantee an additive gap at most} $\log k$ 
for the problem of minimizing the entropy of a joint distribution, with marginals equal to $k$  given input  distributions, for any $k \geq 2.$

In what follows, for the ease of the description, we shall assume that $k= 2^{\kappa}$ for some integer $\kappa \geq 1,$ 
i.e., $k$  is a power of $2$.  A pictorial perspective on the algorithm's behaviour is to imagine that the input distributions are in the 
leaves of a complete binary tree with $k = 2^{\kappa}$ leaves. Each internal node $\nu$ of the tree contains the joint distribution of the 
distributions in the leaves of the subtree rooted at $\nu$. Such a distribution is computed by applying 
the algorithm {\sc  Min-Entropy-Joint-Distr} to the distributions in the children of $\nu$. 

The algorithm builds such a tree starting from the leaves. Thus, 
the  joint distribution of all the input distributions will be given by the distribution computed at the root of the tree. 

Let us first see how in the case of four input distributions, our algorithms does indeed  guarantee that the final matrix is  a
 joint distribution with marginals equal to the input distributions. An high-level  pictorial way to  describe how our algorithm
operates is given in Figure \ref{fig-binarytree}.

\begin{figure}[h]
\centering
\includegraphics[width=5cm]{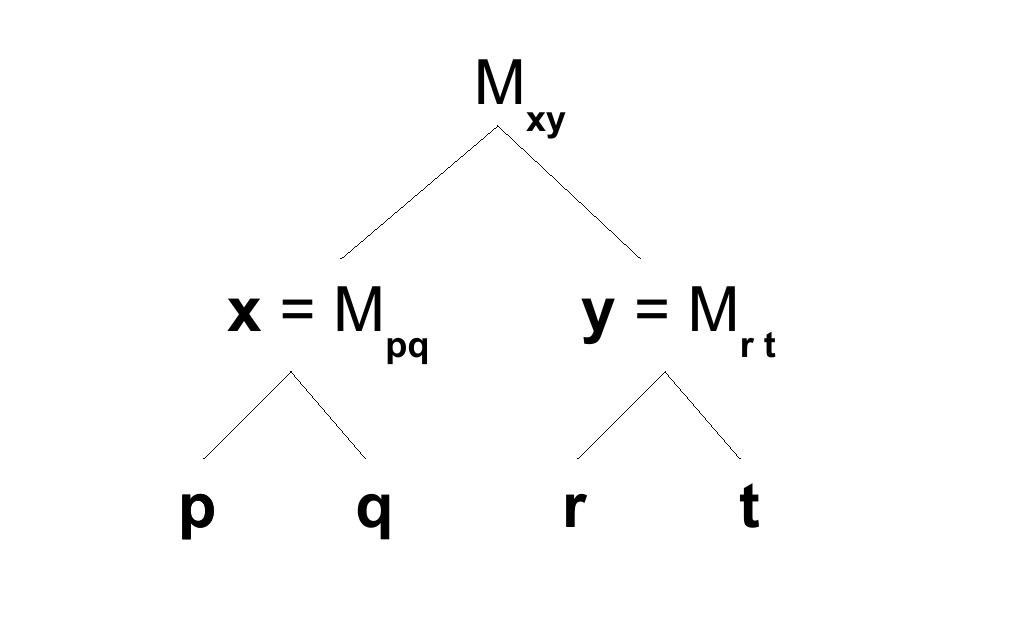}
\caption{The binary tree representing the process of producing the joint distribution for input probability 
vectors $\p,\q,\r,\t.$  \label{fig-binarytree}}
\end{figure}

Given probability distributions $\p$ and $\q$, of dimension $n_{\p}$ and $n_{\q}$, respectively, 
the algorithm  {\sc  Min-Entropy-Joint-Distr} described in Section \ref{sec:approx} produces a matrix
$M_{\p\,\q}$ such that for fixed $\overline{i}$ and $\overline{j}$ it holds that 
\begin{equation} \label{multi-level1}
 \sum_j M_{\p\,\q}[\overline{i},j] = p_{\overline{i}} \qquad \mbox{ and } \qquad \sum_i M_{\p\,\q}[i,\overline{j}] = q_{\overline{j}} .
 \end{equation}

We also have that, for each $i_1 \neq i_2$ the set of entries of $M_{\p\,\q}$ whose sum is $p_{i_1}$ is disjoint from the set of 
entries of $M_{\p\,\q}$ whose sum is $p_{i_2}$. Analogously, 
for each $j_1 \neq j_2$ the set of entries of $M_{\p\,\q}$ whose sum is $q_{j_1}$ is disjoint from the set of 
entries of $M_{\p\,\q}$ whose sum is $q_{j_2}$.

Let us define the probability distribution $\x=(x_1, x_2, \ldots )$ whose components are all and only the non-zero entries of $M_{\p\,\q}$ sorted in 
non-increasing order. Let $n_{\x}$ denote the number of components of $\x$. For each $a= 1, \dots n_{\x}$ let us fix a one-one
mapping $a \leftrightarrow (i,j)$ recording the fact that $x_a = M_{\p\,\q}[i,j]$.

Consider now another pair of distributions $\r, \t$, of dimension $n_{\r}$ and $n_{\t}$, respectively.  
Applying Algorithm 
{\sc  Min-Entropy-Joint-Distr} to $\r, \t$ we obtain  a matrix
$M_{\r\,\t}$ such that for fixed $\overline{i'}$ and $\overline{j'}$ it holds that
$$ \sum_{j'} M_{\r\,\t}[\overline{i'},j'] = p_{\overline{i'}} \qquad \mbox{ and } \qquad \sum_{i'} M_{\r\,\t}[i',\overline{j'}] = q_{\overline{j'}} .$$

As before, for each $i'_1 \neq i'_2$ the set of entries of $M_{\r\,\t}$ whose sum is $p_{i'_1}$ is disjoint from the set of 
entries of $M_{\r\,\t}$ whose sum is $p_{i'_2}$. Also, 
for each $j'_1 \neq j'_2$ the set of entries of $M_{\r\,\t}$ whose sum is $q_{j'_1}$ is disjoint from the set of 
entries of $M_{\r\,\t}$ whose sum is $q_{j'_2}$.

Let $\y=(y_1, y_2, \ldots)$ be the probability distribution whose components are all and only the non-zero entries of $M_{\r\,\t}$ sorted in 
non-increasing order. Let $n_{\y}$ denote the number of components of $\y$. For each $b= 1, \dots n_{\y}$ let us fix a one-one
mapping $b \leftrightarrow (i',j')$ recording the fact that $y_{b} = M_{\r\,\t}[i',j']$.

\medskip
If we now apply algorithm  {\sc  Min-Entropy-Joint-Distr}  on the distributions $\x$ and $\y$ we get a matrix $M_{\x\,\y}$ such that
for fixed $\overline{a}$ and $\overline{b}$ it holds that 
\begin{equation} \label{multi-level2}
 \sum_{b} M_{\x\,\y}[\overline{a},b] = x_{\overline{a}} \qquad \mbox{ and } \qquad \sum_{a} M_{\x\,\y}[a,\overline{b}] = y_{\overline{b}}.
 \end{equation}

\medskip
Let us now define a new $4$-dimensional array $\Mt[i,j,i',j']$ by stipulating that for each $i \in  [n_{\p}],\, j \in  [n_{\q}],\, i' \in  [n_{\r}],\,
j' \in  [n_{\t}],$ the following equalities hold
$$\Mt [i,j,i',j'] = 
\begin{cases}
M_{\x\,\y}[a,b] & \mbox{if there exist } a, b \mbox{ s.t.\ }\\
                & \qquad  a \leftrightarrow (i,j), b \leftrightarrow (i', j')\\
0 & \mbox{otherwise.}
\end{cases}
$$

Then, applying the properties above, for each $\overline{i} \in [n_{\p}],$ we have that 
\begin{align}
\sum_{j \in [n_{\q}]} & \sum_{i' \in [n_{\r}]} \sum_{j' \in [n_{\t}]} \Mt[\overline{i}, j, i', j']\\
                      &=\sum_{\substack{(\overline{i},j) | \\ \exists a, a \leftrightarrow (\overline{i},j)}}  
\sum_{\substack{(i',j') | \\ \exists b, b \leftrightarrow (i',j')}} \Mt[\overline{i}, j, i', j'] \label{multi1}\\
&=
\sum_{a | \exists j, a \leftrightarrow (\overline{i},j)} 
\sum_{b \in [n_{\y}]} M_{\x\,\y}[a, b]   \label{multi2} \\
&=
\sum_{a | \exists j, a \leftrightarrow (\overline{i},j)} 
x_a   \label{multi3} \\
&=
\sum_{j  | M_{\p\,\q}[\overline{i},j] \neq 0}  M_{\p\,\q}[\overline{i}, j] = 
\sum_{j  \in [n_{\q}]}  M_{\p\,\q}[\overline{i}, j]  = p_{\overline{i}}, \label{multi4}
\end{align}
where the equality in (\ref{multi1}) follows by 
restricting the sum over the non-zero entries of $\Mt$;
(\ref{multi2}) follows by the definition of $\Mt$;
(\ref{multi3}) follows by (\ref{multi-level2});
the first part of (\ref{multi4}) follows by the fact that the components of $\x$ 
coincide with non zero entries of $M_{\p\,\q}$;
the first equality in (\ref{multi4}) follows since we are adding to the previous term only entries  
$M_{\p\,\q}[\overline{i}, j]=0$; finally the last equality follows from 
(\ref{multi-level1}).

Proceeding in the same way we can show that for each $\overline{j}, \overline{i'}, \overline{j'}$ we have 
\begin{eqnarray}
\sum_{i \in [n_{\p}]} \sum_{i' \in [n_{\r}]} \sum_{j' \in [n_{\t}]} \Mt[i, \overline{j}, i', j'] &=& q_{\overline{j}} \\
\sum_{i \in [n_{\p}]} \sum_{j \in [n_{\q}]} \sum_{j' \in [n_{\r}]} \Mt[i, j, \overline{i'}, j'] &=& r_{\overline{i'}} \\
\sum_{i \in [n_{\p}]} \sum_{j \in [n_{\q}]} \sum_{i' \in [n_{\r}]} \Mt[i, j, i', \overline{j'}] &=& t_{\overline{j}},
\end{eqnarray}
hence concluding that $\Mt$ is a joint distribution with marginals equal to  $\p, \q, \r, \t,$ as desired.

\medskip
\noindent
{\bf Completing the  argument.} We can now inductively extend the above argument to the case of more
distributions. Assume that we have produced an array 
$\M_{\p^{(1)}, \dots, \p^{(r)}}$  which is a joint distribution with marginals equal to $\p^{(1)}, \dots, \p^{(r)}$, of dimension $n_1, \dots, n_r$ respectively.
Analogously, let us assume that we have produced   
$\M_{\q^{(1)}, \dots, \q^{(s)}}$ which is a joint distribution with marginals equal to $\q^{(1)}, \dots, \q^{(s)},$ of dimension $m_1, \dots, m_s$ respectively. 
This means that for each $\ell=1, \dots, r$ and $1 \leq i \leq n_{\ell}$ and 
for each $\ell'=1, \dots, r$ and $1 \leq j \leq n_{\ell'}$ we have that 
\begin{align*} 
\sum_{i_1, \dots i_{\ell-1}, i_{\ell+1}, \dots, r} &\M_{\p^{(1)}, \dots, \p^{(r)}}[i_1, \dots, i_{\ell-1}, i, i_{\ell+1}, \dots, i_r] \\
                                                   &=p^{(\ell)}_i \\
\sum_{j_1, \dots j_{\ell'-1}, i_{\ell'+1}, \dots, s}&\M_{\q^{(1)}, \dots, \q^{(r)}}[j_1, \dots, j_{\ell'-1}, j, j_{\ell'+1}, \dots, j_s]\\
                                                     &=q^{(\ell')}_j.
\end{align*}

Proceeding as before, let us define 
 the probability distribution $\x$ whose components are all and only the non-zero entries of $M_{\p^{(1)}, \dots, \p^{(r)}}$ sorted in 
non-increasing order. Let $n_{\x}$ denote the number of components of $\x$. 
For each $a= 1, \dots n_{\x}$ let us fix a one-one
mapping $a \leftrightarrow (i_1, \dots, i_r)$ recording the fact that $x_a = M_{\p^{(1)}, \dots, \p^{(r)}}[i_1, \dots, i_r]$.

Let $\y$ be the probability distribution whose components are all and only the non-zero entries of $M_{\q^{(1)}, \dots, \q^{(s)}}$ sorted in 
non-increasing order. Let $n_{\y}$ denote the number of components of $\y$. For each $b= 1, \dots n_{\y}$ let us fix a one-one
mapping $b \leftrightarrow (j_1, \dots, j_s)$ recording the fact that $y_{b} = M_{\q^{(1)}, \dots, \q^{(s)}}[j_1, \dots, j_s]$.

Applying algorithm  {\sc  Min-Entropy-Joint-Distr}  on the distributions $\x$ and $\y$ we get a matrix $M_{\x\,\y}$ such that
for fixed $\overline{k}$ and $\overline{\ell}$
\begin{equation} \label{multi-level2bis}
 \sum_{b} M_{\x\,\y}[\overline{a},b] = x_{\overline{a}} \qquad \mbox{ and } \qquad \sum_{a} M_{\x\,\y}[a,\overline{b}] = y_{\overline{b}}.
 \end{equation}

\medskip
Therefore, we can define a new $r+s$-dimensional array $\Mt[i_1, \dots, i_r, j_1, \dots, j_s]$ 
by stipulating that for each $i_1, \dots, i_r$  such that $i_{\ell} \in [n_k]$ for $\ell=1, \dots, r$  and 
for any $j_1, \dots, j_s$ such that $j_{\ell'} \in [m_{\ell'}]$ for $\ell'=1, \dots, s,$  
$$\Mt [i_1, \dots, i_r, j_1, \dots, j_s] = 
\begin{cases}
M_{\x\,\y}[a,b] & \mbox{if there are $a, b$ s.t.\ }\\
                & a \leftrightarrow (i_1, \dots, i_r),\\
								&\quad b \leftrightarrow (j_1, \dots, j_s)\\
0 & \mbox{otherwise.}
\end{cases}
$$ 

It is  not  hard to see that proceeding like in (\ref{multi1})-(\ref{multi4}) 
one can  show that $\Mt$ is indeed a joint distribution with marginals equal to  $\p^{(1)}, \dots, \p^{(r)}, \q{(1)}, \dots, \q^{(s)}.$

%%%%%%%

\subsection{The pseudocode: Algorithm \ref{algo:multidistribution} }

Algorithm \ref{algo:multidistribution} shows the pseudocode for our procedure.
We  denote by $m^{(i-j)}$ the non-zero components of the distribution that our algorithm builds as joint distribution of 
$\p^{(i)}, \p^{(i+1)}, \dots, \p^{(j)}.$ 

The vector $Ind^{(i-j)}$ is used to record for each component $\m^{(i-j)}[w]$ the indices of the component of the joint probability distribution of $\p^{(i)}, \dots, \p^{(j)}$ which coincides with  $\m^{(i-j)}[w]$.  With respect to the description above $Ind^{(i-j)}$  is used to record the one-one mapping 
between the elements of $\m^{(i-j)}$ and the non-zero elements of the joint distribution of $\p^{(i)}, \p^{(i+1)}, \dots, \p^{(j)}.$
Therefore, in accordance to the above arguments, after the execution of line 17, 
for $w = 1, \dots, |\m^{(i-j)}|$, if, e.g., $Ind^{(i-j)}[w] = \langle s_i[w], s_{i+1}[w], \dots, s_j[w]  \rangle$ it means that
setting  $M^{(i-j)}[s_i[w], s_{i+1}[w], \dots, s_j[w]] \leftarrow \m^{(i-j)}[w]$ and setting the remaining components of $M^{(i-j)}$ to zero, 
the array $M^{(i-j)}$ is a joint distribution matrix for $\p^{(i)}, \dots, \p^{(j)}$ whose non-zero components are equal to the components
of $\m^{(i-j)}.$ Hence, in particular, we have that $H(M^{(i-j)}) = H(\m^{(i-j)}).$

Note that the algorithm explicitly uses this correspondence only for the final array $M^{(1-k)}$ representing the joint distribution of all input distributions. 
Based on the above discussion the correctness of the algorithm can be easily verified.

\subsection{The { additive gap guaranteed by}  {\sc  K-Min-Entropy-Joint-Distribution}}
 In  this section we will prove that the entropy of the 
joint distribution output by the algorithm guarantees { an} additive { gap at most } $\log k.$ 

We will prepare some definitions and lemmas which will be key tools for proving the approximation guarantee of our algorithm. The proof of these technical lemmas is deferred to the next section.

\begin{algorithm}[ht!] 
\small
{\sc  K-Min-Entropy-Joint-Distribution}($\p^{(1)}, \p^{(2)} , \dots, \p^{(k)}$)\\ 
{\bf Input:} prob.\ distributions $\p^{(1)}, \p^{(2)} , \dots, \p^{(k)},$ with $k = 2^{\kappa}$\\
{\bf Output:} A $k$-dimensional array $\mathbf{M} = [m_{i_1, i_2, \dots, i_k}]$ s.t.\ $\sum_{i_1, \dots, i_{j-1}, i_{j+1}, \dots, i_k} 
m_{i_1, \dots, i_{j-1}, t,  i_{j+1}, \dots, i_k} = p^{(j)}_t$ for each $j = 1, \dots, k$ and each $t$.
\begin{algorithmic}[1]
\FOR{$i = 1$ {\bf to} $k$}
\FOR{$j = 1$ {\bf to} $n$}
\STATE{{\bf set} $\m^{(i-i)}[j] = \p^{(i)}_j$ and $Ind^{(i-i)}[j] = \langle j \rangle$}\\ 
~~~~~~~~~~~~~~~~~~\COMMENT{$Ind^{(i-i)}[j]$ is a vector of indices} 
\ENDFOR
\ENDFOR
\STATE{{\bf for} $i=1,\dots, k$  {\bf permute the components of} $\m^{(i-i)}$ and $Ind^{(i-i)}$ using the permutation that sorts $\m^{(i-i)}$ in non-increasing order} 
\FOR{$\ell = 1$ {\bf to} $\kappa$}  \label{k-algomain:1}
\STATE{$i \leftarrow 1, \, j \leftarrow 2^{\ell}$}
\WHILE{$j \leq k$}  
\STATE{$j_1 \leftarrow i+2^{\ell-1}-1, \, j_2 = j_1+1$}
\STATE{$M \leftarrow$ {\sc  Min-Entropy-Joint-Distr}$(\m^{(i-j_1)}, \m^{(j_2-j)})$}   \label{callalgo1}
\STATE{$w \leftarrow 1$}
\FOR{$s=1$ {\bf to} $|\m^{(i-j_1)}|$}
\FOR{$t=1$ {\bf to} $|\m^{(j_2-j)}|$}
\IF{$M[s,t] \neq 0$}
\STATE{$\m^{(i-j)}[w] \leftarrow M[s,t]$}
\STATE{$Ind^{(i-j)}[w]  \leftarrow  Ind^{(i-j_1)}[s] \odot Ind^{(i-j_1)}[t]$}\\ 
~~~~~~~~~~~~~~\COMMENT{$\odot$ denotes the concatenation of vectors} \nonumber
\STATE{$w \leftarrow w+1$}
\ENDIF
\ENDFOR
\ENDFOR
\STATE{{\bf permute the components of} $\m^{(i-j)}$ and $Ind^{(i-j)}$ using the permutation that sorts $\m^{(i-j)}$ in non-increasing order} 
\ENDWHILE 
\STATE{$i \leftarrow j+1, \, j \leftarrow i+2^{\ell}-1$}
\ENDFOR  
\STATE{{\bf set} $M[i_1,i_2,\dots, i_k] = 0$ for each $i_1, i_2, \dots, i_k.$} \label{line:output}
\FOR{$j = 1$ {\bf to} $|\m^{(1-k)}|$}
\STATE{$M[Ind^{(1-k)}[j]] \leftarrow \m^{(1-k)}[j]$}
\ENDFOR
\STATE{{\bf return} $M$}
\end{algorithmic}
\caption{The Min Entropy Joint Distribution Algorithm for $k>2$ distributions}
\label{algo:multidistribution}
\end{algorithm}

Let us define the following:

\begin{definition}
For any $\p = (p_1, \dots, p_n) \in \PP_n$ we denote by $\half(\p)$ the distribution  
$(\frac{p_1}{2}, \frac{p_1}{2}, \frac{p_2}{2}, \frac{p_2}{2}, \dots, \frac{p_n}{2}, \frac{p_n}{2})$ obtained by splitting each component of 
$\p$ into two identical halves.

For any $i \geq 2,$ let us also define $\half^{(i)}(\p) = \half(\half^{(i-1)}(\p)),$ where $\half^{(1)}(\p) = \half(\p)$ and   $\half^{(0)}(\p) = \p.$
\end{definition}

\medskip

We will employ the following two technical lemmas whose proofs are in the next section.

\begin{lemma} \label{monotone-morphism}
For any $\p \preceq \q$ we have also $\half(\p) \preceq \half(\q)$ 
\end{lemma}

\begin{lemma} \label{key-morphism-power}
For any pair of distributions $\p, \q \in \PP_n.$ and any $i \geq 0$, It holds that 
$$\half^{(i)}(\p \wedge \q) \preceq \half^{(i)}(\p) \wedge \half^{(i)}(\q).$$ 
\end{lemma}

\begin{theorem} \label{k-joint:main}
For each $\ell = 0, 1, \dots \kappa$ and $s = 0, 1, 2, \dots, k/2^{\ell}-1$ let $i = i(\ell, s) =  s \cdot 2^{\ell} + 1$ and 
$j = j(\ell, s) = (s+1) \cdot 2^{\ell} = i + 2^{\ell} -1.$ Then, we have 
$$\half^{(\ell)}(\p^{(i)} \wedge \p^{(i+1)} \wedge \cdots \wedge \p^{(j)}) \preceq \m^{(i-j)}.$$
\end{theorem}
\begin{IEEEproof}
The proof is by induction on $\ell.$ The base case follows by definition of the operator $\half^{(\ell)}$ and the 
fact that the algorithm sets $\m^{(i-i)} = \p^{(i)},$ for each $i$ hence in particular  $\m^{(i-i)} = \p^{(i)} = \half^{(0)}(\p^{(i)}),$ which 
proves the desired inequality.

We now prove the induction step. Let $\ell > 0.$ It is enough to consider only the case $s = 0,$ since the other cases are perfectly analogous. 

Therefore, $i = 1$ and $j = 2^{\ell}.$ Using the notation employed in the pseudocode, let $j_1 = 2^{\ell-1}, \, j_2 = 2^{\ell-1}+1.$ 
By induction hypothesis we can assume that 
\begin{equation} \label{eq:inductionhypothesis-1}
\half^{(\ell-1)}(\p^{(i)} \wedge \p^{(i+1)} \wedge \cdots \wedge \p^{(j_1)}) \preceq \m^{(i-j_1)}
\end{equation}
\begin{equation} \label{eq:inductionhypothesis-2}
\half^{(\ell-1)}(\p^{(j_2)} \wedge \p^{(j_2+1)} \wedge \cdots \wedge \p^{(j)}) \preceq \m^{(j_2-j)}.
\end{equation}

It follows that
\begin{eqnarray}
\phantom{a}& &\hspace{-0.8cm} \half^{(\ell)}\left(\bigwedge_{\iota = i}^j \p^{(\iota)}  \right) = \nonumber \\
&=&   \half^{(\ell)}\left( \left( \bigwedge_{\iota = i}^{j_1} \p^{(\iota)} \right) 
                               \wedge \left( \bigwedge_{\iota = j_2}^j \p^{(\iota)}    \right)  \right)   \label{eq:half1}\\
    &=& \half \left( \half^{(\ell-1)}\left( \left( \bigwedge_{\iota = i}^{j_1} \p^{(\iota)} \right) 
                                                          \wedge \left( \bigwedge_{\iota = j_2}^j \p^{(\iota)}    \right)  \right) \right) \label{eq:half2}\\
    &\preceq& \half\left( \half^{(\ell-1)} \left( \bigwedge_{\iota = i}^{j_1} \p^{(\iota)}  \right)    \wedge
                             \half^{(\ell-1)} \left( \bigwedge_{\iota = j_2}^{j} \p^{(\iota)}  \right)   \right) \nonumber \\
                             & &   \label{eq:half3}\\
    &\preceq& \half\left( \m^{(i-j_1)} \wedge \m^{(j_2-j)} \right) \label{eq:half4}\\
    &\preceq& \m^{(i-j)}  \label{eq:half5}
\end{eqnarray}
where 
\begin{itemize}
\item (\ref{eq:half2}) follows from (\ref{eq:half1}) by the definition of the operator $\half$;
\item (\ref{eq:half3}) follows from (\ref{eq:half2}) by Lemma \ref{key-morphism-power};
\item (\ref{eq:half4}) follows from (\ref{eq:half3}) by the induction hypotheses (\ref{eq:inductionhypothesis-1})-(\ref{eq:inductionhypothesis-2}) ;
\item (\ref{eq:half5}) follows from (\ref{eq:half4}) by observing that the components of $\m^{(i-j)}$ 
coincide with the components of the array $M$ output by algorithm 
{\sc Min-Entropy-Joint-Distribution} executed on the distributions $\m^{(i-j_1)}$ and $\m^{(j_2-j)}.$ 
Let $\z = \m^{(i-j_1)} \wedge \m^{(j_2-j)}$ and $|\z|$ denote the number of components of $\z.$
By the analysis presented in the previous section we have that we can partition the components of $M$ (equivalently, the components
of $\m^{(i-j)}$) into subsets 
$M_1, M_2, \dots, M_{|\z|}$ such that 
\begin{itemize}
\item $1 \leq |M_i| \leq 2$
\item for each $i = 1, \dots, |\z|,$ it holds that  $\sum_{x \in M_i} x = z_i$;
\end{itemize}
Therefore---assuming, w.l.o.g., that the components of $\m^{(i-j)}$ are reordered such that those in $M_i$ immediately precede those in 
$M_{i+1}$---we have $\half(\z) = \m^{(i-j)} P$ where $P = [p_{i\, j}]$ is a doubly stochastic matrix defined by 
$$p_{i\,j} = \begin{cases}
\frac{1}{2} &  \mbox{if ($i$ is odd {\bf and} } j \in \{i, i+1\} \mbox{)} \\
            & \ \mbox{\bf or ($i$ is even {\bf and} } j \in \{i, i-1\}); \\
0 & otherwise
\end{cases} 
$$
\end{itemize}
from which it follows that $\half(\z) \preceq \m^{(i-j)}$ yielding (\ref{eq:half5}).
\end{IEEEproof}

\smallskip
An immediate consequence of the last theorem is the following
\begin{corollary}
For any  $k$ probability distributions $\p^{(1)}, \dots, \p^{(k)}$ let $M$ be the joint distribution, with marginals equal to  $\p^{(1)}, \dots, \p^{(k)}$,
 output by algorithm 
{\sc K-Min-Entropy-Joint-Distribution}. Then,  
$$H(M) \leq H(\p^{(1)} \wedge \p^{(2)} \wedge \cdots \p^{(k)}) + \lceil \log k \rceil$$ 
\end{corollary}
\begin{IEEEproof}
Let $k$ be a power of $2$. Otherwise, one can duplicate  some of the probability distributions until there are  $k' = 2^{\lceil \log k \rceil}$ of them. 
By Theorem \ref{k-joint:main} we have 

\begin{align*}
\half^{(\lceil \log k \rceil )}&(\p^{(1)} \wedge \p^{(2)} \wedge \cdots \p^{(k)})\\
                               &=  
\half^{(\log k'  )}(\p^{(1)} \wedge \p^{(2)} \wedge \cdots \p^{(k')})  \preceq \m^{(1-k)}.
\end{align*}

Therefore, by the Schur-concavity of the entropy we have 
\begin{align*}
H(M) &= H(\m^{(1-k)})\\
     &\leq H(\half^{(\lceil \log k \rceil )}(\p^{(1)} \wedge \p^{(2)} \wedge \cdots \wedge\p^{(k)})) \\
     &= H(\p^{(1)} \wedge \p^{(2)} \wedge \cdots \wedge\p^{(k)}) + \lceil \log k \rceil,
\end{align*}
where the last equality follows by the simple observation that 
for any probability distribution $\x$ and integer $i \geq 0$ we have $H(\half^{(i)}(\x)) = H(\x) + i.$
\end{IEEEproof}

\medskip

We also have the following lower bound which, together with the previous corollary implies that 
our algorithm guarantees an additive { gap of at most} $\log k$ { bits} 
 for the problem of 
computing the joint distribution of minimum entropy of $k$ input distributions. 

\begin{lemma}\label{lowerboundx}
Fix $k$ distributions $\p^{(1)}, \p^{(2)},  \cdots,  \p^{(k)}$. For any $M$ being a joint distribution with marginals  
$\p^{(1)}, \p^{(2)},  \cdots, \p^{(k)}$, it holds that 
$${H(M) \geq H(\p^{(1)} \wedge \p^{(2)} \wedge \cdots \wedge \p^{(k)})}.$$
\end{lemma}
\begin{IEEEproof}
For each $i = 1, \dots, k,$ the distribution $\p^{(I)}$ is an aggregation of $M$, hence $M \preceq \p^{(i)}.$

By definition of the greatest lower bound operator $\wedge$ for any distribution $\x$ such that for each $i$ it holds that $\x \prec \p^{(i)}$ we have 
$\x \preceq \p^{(1)} \wedge \p^{(2)} \wedge \cdots \p^{(k)}$. Therefore, in particular we have
$M \preceq \p^{(1)} \wedge \p^{(2)} \wedge \cdots \p^{(k)},$ which, by the Schur concavity of the entropy 
gives the desired result. 
\end{IEEEproof}

\begin{remark}
The time complexity of Algorithm \ref{algo:multidistribution} is dominated by the time to build the output matrix in line \ref{line:output}, 
which takes $O(n^k).$
However, if the output matrix is returned in sparse form and in line \ref{callalgo1}
the improved implementation of algorithm {\sc Min-Entropy-Joint-Distribution} is used (see section \ref{section:improvedtime} 
and the appendix),  the time for the overall construction
is upper bounded by
$\sum_{\ell =1}^{\log k} O(\frac{k}{2^{\ell}} 2^{\ell-1} n \log(2^{\ell-1} n)) = O(n k  \log (n k)).$
To see this, observe that the main {\bf for} loop in line \ref{k-algomain:1} is executed $O(\log k)$ times and 
in each iteration $\ell=1, \dots, \log k$ there are $\frac{k}{2^{\ell}}$ executions of {\sc Min-Entropy-Joint-Distribution}
over distributions having $O(2^{\ell-1} n)$ non-zero entries and the algorithm employs the arrays $Ind$ in order to 
perform the computation only considering the non-zero entries of these distributions. 
\end{remark}
   
 Summarising we have shown the following
 
 \begin{theorem}\label{th:more}
Let $\p^{(1)}, \dots, \p^{(m)} \in \PP_n$. Let $M^*$ be a joint distribution with marginals  $\p^{(1)}, \dots, \p^{(m)}$ of \emph{minimum} entropy among all 
the joint distribution having marginals equal to  $\p^{(1)}, \dots, \p^{(m)}.$
Let $M$ be the joint distribution of $\p^{(1)}, \dots, \p^{(m)}$ output by our algorithm. Then, 
$$H(M) \leq H(M^*) + \lceil \log(m) \rceil.$$
Hence, our (polynomial time) algorithm { guarantees} an additive { gap of } $\log(m).$
\end{theorem}

Using { Lemma \ref{lowerboundx} and Theorem \ref{th:more} above},  and Theorem 3 of \cite{ko1}, we obtain the 
following version of the functional representation lemma (please see  the discussion in Section  \ref{sub:fl}
of the present paper).

\begin{corollary}\label{cor:frl}
Let $X$ and $Y$ be two arbitrary random variables with joint distribution $p(x,y)$, where $X$ takes 
values $x_1, \ldots , x_k$. Let $\p^{(1)}, \dots, \p^{(k)}$ be the
distribution of the conditioned r.v. $Y|X=x_1, \ldots , Y|X=x_k$, respectively. Then, for any r.v. $Z$ independent from $X$ for which 
there exist a  function $f$
such that  $(X,Y)=(X,f(X,Z))$,  it holds that
$$H(Z)\geq H(\p^{(1)} \wedge \p^{(2)} \wedge \cdots \wedge \p^{(k)}).$$ 
Conversely, there exists a $Z$ independent from $X$ and a function $f$
for which $(X,Y)=(X,f(X,Z))$ such that 
$$H(Z)\leq H(\p^{(1)} \wedge \p^{(2)} \wedge \cdots \wedge\p^{(k)})+ \log k.$$
\end{corollary}
  
\subsection{The proofs of the two technical lemmas about the operator $\half$}

\medskip
\noindent
{\bf Lemma \ref{monotone-morphism}.}
{\em For any $\p \preceq \q$ we have also $\half(\p) \preceq \half(\q)$.} 

\begin{IEEEproof}
It is easy to see that assuming $\p$ and $\q$ rearranged in order to have 
$p_1 \geq p_2 \geq \dots \geq p_n$ and $q_1 \geq q_2 \geq \dots \geq q_n$ we also have
$\half(\p)_1 \geq \half(\p)_2 \geq \dots \geq \half(\p)_{2n}$ and $\half(\q)_1 \geq \half(\q)_2 \geq \dots \geq \half(\q)_{2n}.$

By assumption we also have that for each $j = 1, \dots, n$ it holds that $\sum_{i=1}^j p_i \leq  \sum_{i=1}^j p_i.$

Therefore, for each $j=1, \dots 2n$ it holds that 
\begin{align*}
\sum_{i=1}^j \half(\p)_i &= \frac{1}{2} \sum_{i=1}^{\lceil j/2 \rceil} p_i + \frac{1}{2} \sum_{i=1}^{\lfloor j/2 \rfloor} p_i \\
                         &\leq
\frac{1}{2} \sum_{i=1}^{\lceil j/2 \rceil} q_i + \frac{1}{2}\sum_{i=1}^{\lfloor j/2 \rfloor} q_i  = \sum_{i=1}^j \half(\q)_i,
\end{align*}
proving that $\half(\p) \preceq \half(\q).$
\end{IEEEproof}

\medskip

\begin{fact} \label{key-morphism}
For any pair of distributions $\p, \q \in \PP_n.$ It holds that 
$$\half(\p \wedge \q) \preceq \half(\p) \wedge \half(\q).$$ 
\end{fact}
\begin{IEEEproof}
By   Lemma \ref{monotone-morphism} we have that 
$$\half(\p \wedge \q) \preceq \half(\p) \qquad \mbox{and} \qquad \half(\p \wedge \q) \preceq \half(\q)$$
Then, by the property of the operator $\wedge$ which gives the greatest lower bound we have the desired result.
\end{IEEEproof}

On the basis of Fact \ref{key-morphism} we can extend the result to "powers" of the operator $\half$ and have 
our Lemma \ref{key-morphism-power}. 

\medskip
\noindent
{\bf Lemma \ref{key-morphism-power}.} {\em 
For any pair of distributions $\p, \q \in \PP_n.$ and any $i \geq 0$, It holds that 
$$\half^{(i)}(\p \wedge \q) \preceq \half^{(i)}(\p) \wedge \half^{(i)}(\q).$$ 
}
\begin{IEEEproof}
We argue by induction on $i$. The base case $i=1$ is given by the previous Fact \ref{key-morphism}. Then,
for any $i > 1$ 
\begin{align*}
\half^{(i)}(\p \wedge \q) &= \half(\half^{(i-1)}(\p \wedge \q)) \\
                          &\preceq \half(\half^{(i-1)}(\p)  \wedge \half^{(i-1)}(\q)) \\
													&\preceq \half(\half^{(i-1)}(\p)) \wedge  \half(\half^{(i-1)}(\p))
\end{align*}
from which the desired result immediately follows. The first $\preceq$-inequality follows by induction hypothesis and the
second inequality by Fact \ref{key-morphism}.
\end{IEEEproof}

\section{Conclusions}
In this paper we have studied the problem of finding a minimum entropy joint distribution with \emph{fixed} marginals.
We have pointed out that this problem naturally arises in a  variety of situations: causal inference, 
one-shot channel simulation, metric computation for dimension reduction, optimization in the transportation polytope,
and several others. Our main result consists in a polynomial time algorithm to find an $\M\in\CC(\p,\q)$ such 
that $H(\M)\leq OPT+1$ bit, where $OPT=\min_{\mathbf{N}\in \CC(\p, \q)} H(\mathbf{N}).$ 
We ave also shown that our approach (relying on majorization among probability distributions)
allows us to easily extend our results to R\'{e}nyi entropies of arbitrary positive orders 
(thus generalizing the result for the Shannon entropy where the latter is equal to the R\'{e}nyi entropy of order 1).

There are many possible extensions of our work.
Firstly, although our result for the minimum entropy bivariate joint distribution with {fixed} two marginals
seems quite tight, it is very likely that  a more direct approach (i.e., that does not rely on the iterative construction of
Section \ref{sec:more}) could give better results for multivariate joint distributions.
Another interesting problem would be to extend our results to the case in which one seeks a minimum entropy 
bivariate joint distribution with  marginals ``close'' to given ones, for appropriate measures of closeness.
Finally,  a natural research  problem is related to the scenario considered in Section \ref{sub:fl}: 
Given arbitrary correlated r.v.'s $X$ and $Y$, it would be interesting
to find a r.v. $Z$, independent from $X$, such that  the 
pair of 
r.v.'s $(X,f(X,Z))$ is distributed like  $(X,Y)$, 
for appropriate deterministic function 
$f$, for which  \emph{both} $H(Z)$ and $H(Y|Z)$ are close to their lower bounds.

\section*{Acknowledgments}
The authors want to thank  Executive Editor Professor I. Sason, Associate Editor Professor I. Kontoyiannis,  and the anonymous referees 
for  many useful comments and  suggestions.

\appendix
\begin{algorithm*}%[ht!]
\small 
{\sc  Min-Entropy-Joint-Distribution-Sparse}($\p , \q$)\\ 
{\bf Input:} prob.\ distributions $\p = (p_1, \dots, p_n)$ and $\q= (q_1, \dots, q_n)$\\
{\bf Output:} A Coupling  $\mathbf{M} = [m_{i\,j}]$ of $\p$ and $\q$  in sparse representation  
 $\mathbf{L} = \{\left(m_{i\,j}, (i,j)\right) \mid m_{i\,j} \neq 0\}$\\
\begin{algorithmic}[1]
\STATE{{\bf if} $\p \neq \q$, let $i = \max \{j \mid p_j \neq q_j\}$; 
{\bf if} $p_i < q_i$ {\bf then swap} $\p \leftrightarrow \q$} \label{newalgosparse:initiate-swap}
\STATE{$\z = (z_1, \dots, z_n) \leftarrow \p \land \q, \, \,  \, {\bf L} \leftarrow \emptyset$}
\STATE{{\sc CreatePriorityQueue}(${\cal Q}^{(row)}$), \,  $qrowsum \leftarrow 0$} 
\STATE{{\sc CreatePriorityQueue}(${\cal Q}^{(col)}$), \,  $qcolsum \leftarrow 0$}
\FOR{$i =n$ {\bf downto} $1$}
\STATE{$z^{(d)}_i \leftarrow z_i, \, \, z^{(r)}_i \leftarrow 0$}
\IF{$qcolsum + z_i  > q_i$}  \label{newalgosparse:if1-start}
\STATE{$(z_{i}^{(d)},  z_{i}^{(r)}, I, qcolsum) \leftarrow  
\mbox{\sc  Lemma\ref{lemma:pezzettini}-Sparse}(z_i, q_i, {\cal Q}^{(col)}, qcolsum)$ \label{newalgosparse:if1-1}}   
\STATE{{\bf for each} $(m, \ell) \in I$ {\bf do}  
    ${\bf L} \leftarrow {\bf L} \cup\{(m, (\ell,  i)\} $} \label{sparsemove-1-1}
\STATE{{\bf if} $z^{(r)}_i > 0$ {\bf then } {\sc Insert}(${\cal Q}, (z^{(r)}_i, i)$); \, $qcolsum \leftarrow qcolsum + z^{(r)}_i$} 
\ELSE[$qcolsum + z_i  = q_i$]
\WHILE{${\cal Q}^{(col)} \neq \emptyset$ }
\STATE{$(m, \ell) \leftarrow \mbox{\sc ExtractMin}({\cal Q}^{(col)}$), $qcolsum \leftarrow qcolsum - m,
\, {\bf L} \leftarrow {\bf L} \cup \{(m, (\ell, i))\}$}
\ENDWHILE
\ENDIF 
\IF{$qrowsum + z_i > p_i$}   \label{newalgosparse:if2-start}
\STATE{($z_{i}^{(d)},  z_{i}^{(r)}, I, qrowsum$) $\leftarrow$  \label{newalgosparse:if2-1}
{\sc  Lemma\ref{lemma:pezzettini}-Sparse}($z_i, p_i, {\cal Q}^{(row)}, qrowsum$)} 
\STATE{{\bf for each} $(m, \ell) \in I$ {\bf do}
    ${\bf L} \leftarrow {\bf L} \cup\{(m, (i, \ell)\} $} \label{sparsemove-2-1}
\STATE{{\bf if} $z^{(r)}_i > 0$ {\bf then } {\sc Insert}(${\cal Q}^{(row)}, (z^{(r)}_i, i)$); \, 
$qrowsum \leftarrow qrowsum + z^{(r)}_i$} 
\ELSE[$qrowsum + z_i  = p_i$]
\WHILE{${\cal Q}^{(row)} \neq \emptyset$ }
\STATE{$(m, \ell) \leftarrow \mbox{\sc ExtractMin}({\cal Q}^{(row)}$), $qrowsum \leftarrow qrowsum - m,
\, {\bf L} \leftarrow {\bf L} \cup \{(m, (i, \ell))\}$}
\ENDWHILE
\ENDIF 
\STATE{${\bf L} \leftarrow {\bf L} \cup\{(z^{(d)}_i, (i,i))\};$}
\ENDFOR
\end{algorithmic}
\caption{The Min Entropy Joint Distribution Algorithm outputting a sparse representation of ${\bf M}$}
\label{algo:Algonewnewsparse}
\end{algorithm*}
\begin{algorithm*}[ht!]
\small
{\sc  Lemma\ref{lemma:pezzettini}-Sparse}($z, x, {\cal Q}, qsum$)\\ 
{\bf Input:} reals $z > 0, \, x \geq 0,$ {  and priority queue ${\cal Q}$  
s.t.  $\left(\sum_{(m, \ell) \in {\cal Q}} m\right) = qsum$ and $qsum + x \geq z$ } \\
{\bf Output:} $z^{(d)}, z^{(r)} \geq 0,$ and $I \subseteq {\cal Q}$ s.t. $z^{(d)} + z^{(r)} = z,$ and $z^{(d)} + 
\sum_{(m, \ell) \in I} m = x.$
\begin{algorithmic}[1]
\STATE{$I \leftarrow \emptyset, \, sum \leftarrow 0$}  
\WHILE{${\cal Q} \neq \emptyset$ {\bf and} $sum + \mbox{\sc Min}({\cal Q}) < x$}
\STATE{$(m, \ell) \leftarrow \mbox{\sc ExtractMin}({\cal Q}), \, qsum \leftarrow qsum-m$} 
\STATE{$I \leftarrow I \cup \{(m, \ell)\}, \, sum \leftarrow sum + m$} 
\ENDWHILE
\STATE{$z^{(d)} \leftarrow x - sum, \, z^{(r)} \leftarrow z - z^{(d)}$}
\STATE{{\bf return} ($z^{(d)}, z^{(r)}, I, qsum$)}
\end{algorithmic}
\caption{The procedure implementing Lemma \ref{lemma:pezzettini} for the sparse implementation}
\label{algo:Lemma-sparse}
\end{algorithm*}

\newpage

\end{document}